\documentclass[12pt, a4paper]{article}
\usepackage[height=25cm,width=16cm]{geometry}
\usepackage{feynmp}
\usepackage{amsmath}
\usepackage{ascmac}
\usepackage{dcolumn}
\usepackage{bm,here}
\usepackage{subfig}
\usepackage{comment}
\usepackage{ifpdf}
\usepackage{slashed}
\usepackage{colortbl}
\usepackage{color}
\usepackage{comment}
\usepackage{braket}
\usepackage[mathscr]{eucal}
\usepackage[sort&compress, numbers, merge]{natbib}

\usepackage{cancel}

\ifpdf
\usepackage{graphicx}     
\usepackage[bookmarksopen,colorlinks=true,linkcolor=bblue,citecolor=ppink,urlcolor=ppink]{hyperref}
\else     
\usepackage[dvipdfmx]{graphicx}     
\usepackage[dvipdfmx,bookmarksopen,colorlinks=true,linkcolor=bblue,citecolor=ppink,urlcolor=ppink]{hyperref}
\fi

\usepackage{multicol}
\definecolor{red}{rgb}{1,0,0}
\definecolor{ppink}{rgb}{0.921545,0.440586,0.687243}
\definecolor{bblue}{rgb}{0.400000,0.400000,1.000000}
\usepackage[charter]{mathdesign}
\usepackage{soul}
\usepackage{wrapfig}

\newcommand{\prn}[1]{\left( {#1} \right)}

\newcommand{\der}{\partial}

\newcommand{\abs}[1]{\left\vert {#1} \right\vert}

\newcommand\blfootnote[1]{%
	\begingroup
	\renewcommand\thefootnote{}\footnote{#1}%
	\addtocounter{footnote}{-1}%
	\endgroup
}

\begin{document}
	
\begin{titlepage}

\begin{flushright}
    
\end{flushright}

\begin{center}

	\vskip 1.5cm
	{\Large \bf Thermal Real Scalar Triplet Dark Matter}

	\vskip 2.0cm
	{\large Taisuke Katayose$^{1, *}$\blfootnote{$^*$taisuke.katayose@het.phys.sci.osaka-u.ac.jp},
			Shigeki Matsumoto$^{2, \dagger}$\blfootnote{$^\dagger$shigeki.matsumoto@ipmu.jp}, \\ [.3em]
			Satoshi Shirai$^{2, \ddagger}$\blfootnote{$^\ddagger$satoshi.shirai@ipmu.jp}
			and
			Yu Watanabe$^{2, \$}$\blfootnote{$^{\$}$yu.watanabe@ipmu.jp} }
			
	\vskip 2.0cm
	$^1${\sl Department of Physics, Osaka University, Toyonaka, Osaka 560-0043, Japan} \\ [.3em]
	$^2${\sl Kavli IPMU (WPI), UTIAS, University of Tokyo, Kashiwa, Chiba  277-8583, Japan} \\ [.3em]
	
    \vskip 3.5cm
    \begin{abstract}
        \noindent
        Real scalar triplet dark matter, which is known to be an attractive candidate for a thermal WIMP, is comprehensively studied paying particular attention to the Sommerfeld effect on the dark matter annihilation caused by the weak interaction and the other interaction between the dark matter and the Higgs boson. We find a parameter region that includes the so-called 'WIMP-Miracle' one is still surviving, i.e. it respects all constraints imposed by dark matter searches at collider experiments, underground experiments (direct detection) and astrophysical observations (indirect detection). The region is also found to be efficiently searched for by various near future experiments. In particular, the XENONnT experiment will cover almost the entire parameter region.
    \end{abstract}
			
\end{center}
		
\end{titlepage}
	
\tableofcontents
\newpage
\setcounter{page}{1}
	
\section{Introduction}
\label{sec: intro}

Revealing the identity of dark matter in our universe is one of the most important problems in particle physics. Among various candidates, the thermal dark matter, i.e. the one that experienced the thermal equilibrium with standard model (SM) particles in the early universe, is known to be an attractive candidate, as it is free from the the initial condition problem of the dark matter density and possible to be detected based on the interaction dependable on maintaining the above equilibrium. Moreover, the thermal dark matter with electroweak-sized interaction and mass, which is called the weakly interacting massive particle (WIMP), is particularly interesting, as it could naturally explain the observed dark matter density by the so-called freeze-out mechanism, i.e. the one makes the modern cosmology successful via the Big Bang nucleosynthesis and the recombination of the universe. The fact that the WIMP naturally explains the observed dark matter density is called the `WIMP-Miracle'.

With these attractive features mentioned above, many types of WIMPs are indeed being intensively searched for in various experiments and observations. Uncharted types of WIMPs are, however, still exist, and one of them is the WIMP that is charged under the weak interaction, i.e. the one described by an electrically neutral component in the $\mathrm{SU(2)_L}$ multiplet of either a scalar, fermion, or vector field\,\cite{Cirelli:2005uq}. Such a WIMP is sometimes called the electroweakly interacting massive particle (EWIMP). The EWIMP is also known to be an attractive WIMP candidate from the viewpoint of minimality, as it is described by a smaller number of new physics parameters than other WIMPs, and the `WIMP-Miracle', as its interaction is nothing but the weak interaction itself. Moreover, the EWIMP is predicted in various attractive new physics scenarios; the Wino (Majorana fermion triplet EWIMP) or Higgsino (pseudo Dirac fermion doublet EWIMP) in the minimal supersymmetric SM (MSSM) and the minimal dark matter (Majorana fermion quintplet EWIMP, etc.)\,\cite{Cirelli:2009uv}. Among various EWIMPs, we comprehensively study the EWIMP that is one of the most minimal candidates while not very much studied compared to others, namely the real scalar triplet EWIMP.

There are several (recent) studies on the real scalar triplet dark matter. In Ref.\,\cite{Cirelli:2007xd}, the relic abundance of the dark matter has been calculated including the Sommerfeld effect caused by the weak interaction. In Ref.\,\cite{Chiang:2020rcv}, the collider phenomenology of the dark matter has been thoroughly studied taking into account the above relic abundance but also with a rough estimate on the effect of the scalar interaction between the dark matter and the Higgs boson. In Ref.\,\cite{Arakawa:2021vih}, the phenomenology of the dark matter including calculations of signals at colliders, underground laboratories and astrophysical observatories have been comprehensively studied regardless of the `WIMP-Miracle' condition, i.e. focusing mainly on the case that the dark matter has an electroweak scale mass rather than the TeV scale mass. In this paper, we comprehensively study the dark matter by calculating the relic abundance including the Sommerfeld effect with the full treatment of the scalar interaction, the direct dark matter detection signal at two-loop level and the indirect dark matter detection signal including the Sommerfeld effect with the scalar interaction. It is found that taking into account both the Sommerfeld effect and the scalar interaction allows us to find the `WIMP-Miracle' region of the real scalar triplet dark matter. Together with the scattering cross section between the dark matter and a nucleon at two-loop level, those calculations are also mandatory to quantitatively discuss the future prospects of detecting the dark matter.

This paper is organized as follows. In Section\,\ref{sec: model}, we give the lagrangian describing the real scalar triplet dark matter. We also calculate the favored parameter region, in which the theory does not break down up to high energy scale, by solving appropriate renormalization group equations with two-loop $\beta$ functions explicitly shown in appendix\,\ref{appendix: beta function}. In Section\,\ref{sec: abundance}, we calculate the relic abundance including the Sommerfeld effect. In Section\,\ref{sec: detection}, we consider various constraints on the dark matter from collider, direct and indirect detection experiments at present and in near future. Finally, we summarize our discussion in Section\,\ref{sec: conclusion}.

\section{Real scalar triplet dark matter}
\label{sec: model}

After giving the minimal theory for the real scalar triplet dark matter and addressing interactions between the dark matter and SM particles, we discuss the preferred range of coupling constants for the interactions when we postulate that the theory does not break down up to high-energy scale by solving corresponding renormalization group equations (RGEs).

\subsection{Lagrangian}
\label{subsec: Lagrangian}

The minimal and renormalizable theory describing interactions between the real scalar triplet field $\chi$ and the SM particles is defined by the following simple lagrangian:
\begin{equation}
    \label{lagrangian}
    \mathcal{L}
    = \mathcal{L}_{\rm SM}
    + \frac{1}{2} (\abs{D_\mu \chi}^2 - \mu_\chi^2\abs{\chi}^2)
    - \lambda_{\chi H} \abs{H_{\rm SM}}^2 \abs{\chi}^2
    - \lambda_{\chi} \abs{\chi}^4, 
\end{equation}
where $\mathcal{L}_{\rm SM}$ is the SM lagrangian, $D_\mu = \partial_\mu + i g_2 T^a W^a_\mu$ is the covariant derivative acting on the triplet field $\chi$ with $g_2$, $T^a$ and $W^a_\mu$ being the coupling constant, the generator and the gauge boson field of the weak interaction, respectively, while $H_{\rm SM}$ is the SM Higgs doublet field. The triplet scalar field $\chi$ has three components as $(\chi^+, i\chi^0, \chi^-)^T$ with $\chi^0$ being its neutral component and describing the dark matter. In order to make the dark matter stable, a $\mathbb{Z}_2$ symmetry is imposed, where $\chi$ (SM particles) is odd (even) under the symmetry.\footnote{Without the $\mathbb{Z}_{2}$ symmetry, the term $H_{\rm SM}^\dag \sigma^a H_{\rm SM}\chi_a$ exists and induces the decay of $\chi^0 \to h h$ at tree level.}

After the electroweak symmetry breaking, $H_{\rm SM} = (0, v_H + h)^T/\sqrt{2}$ with $v_H \simeq 246$\,GeV being the vacuum expectation value of the Higgs field and $h$ being the field describing the physical mode of the field (Higgs boson), the above lagrangian is expanded as follows: 
\begin{equation}
    \label{lagrangianEWSB}
    \mathcal{L}
    = \mathcal{L}_{\rm SM}
    - \frac{1}{2} \chi^0 (\partial^2 + m_\chi^2) \chi^0
    - \chi^+ (\partial^2 + m_\chi^2) \chi^-
    - \lambda_\chi [(\chi^0)^2 + 2 \chi^+ \chi^-]^2 + \mathcal{L}_{\rm int},
\end{equation}
where $m_\chi^2 = (\mu_\chi^2 + \lambda_{\chi H} v_H^2)$, and $\mathcal{L}_{\rm int}$ involves all interactions between the dark matter $\chi^0$ (as well as its SU(2)$_L$ partners $\chi^\pm$) and the SM particles. Here, we take the unitary gauge to write down the interactions. Explicit forms of the interactions in $\mathcal{L}_{\rm int}$ are given by
\begin{align}
    \mathcal{L}_{\mathrm{int}}
    &= g_2 [ W^\mu (\chi^0 \overleftrightarrow{\der_\mu} \chi^+) + h.c.] - i g_2 (c_W Z^\mu + s_W A^\mu) (\chi^+ \overleftrightarrow{\partial_\mu} \chi^- )
    \nonumber \\
    &+ (g_2^2/2) (W_\mu W^\mu \chi^+ \chi^+ + h.c.) + g_2^2 W_\mu W^{\mu \dagger} [\chi^+ \chi^- + (\chi^0)^2]
    \nonumber \\
    &- g_2^2 (c_W Z^\mu + s_W A^\mu) (i W_\mu \chi^0 \chi^+ + h.c.) + g_2^2 (c_W Z^\mu + s_W A^\mu)^2 \chi^+ \chi^-
    \nonumber \\
    &- \lambda_{\chi H} v_H h [(\chi^0)^2 + 2 \chi^+ \chi^-] - (\lambda_{\chi H}/2)\,h^2 [(\chi^0)^2 + 2 \chi^+ \chi^-],
\end{align}
where $c_W\,(s_W) = \sin \theta_W\,(\cos \theta_W)$ with $\theta_W$ being the Weinberg angle. The dark matter $\chi^0$ as well as its SU(2)$_L$ partners $\chi^\pm$ interact with SM particles through the weak interaction and the scalar interaction between $H_{\rm SM}$ and $\chi$, so that the strength of the interactions is governed by the couplings $g_2$ and $\lambda_{\chi H}$. Since the dark matter does not couple to the $Z$ boson, it does not suffer from too much severe constraints from the direct dark matter detection.

As seen in eq.(\ref{lagrangianEWSB}), the neutral and charged components of the triplet scalar field $\chi$ are degenerate in mass at tree level. On the other hand, at one-loop level, radiative corrections break the degeneracy. When the mass of the triplet field is much larger than the vacuum expectation value $v_H$, the mass difference between the components is estimated to be\,\cite{Cirelli:2005uq}:
\begin{equation}
    \label{eq: mass difference}
    \delta m \equiv m_{\chi^\pm} - m_{\chi^0} \simeq g_2^2/(4\pi)\,(g_2 v_H/2)\,\sin^2 (\theta_W/2) \simeq 166\,\mathrm{MeV}.
\end{equation}
The mass difference is proportional to a single power of the vacuum expectation value due to the nature (threshold singularity) of the corrections. Since the mass of the triplet dark matter is expected to be ${\cal O}(1)$\,TeV as discussed in the next section, the above estimation indicates that the dark matter is highly degenerate ($\lesssim 10^{-4}$) with its SU(2)$_L$ partners.

\subsection{Preferred region of model parameters}
\label{subsec: Landau pole}

As seen from the lagrangian in eq.\,(\ref{lagrangianEWSB}), physics of the triplet dark matter and its SU(2)$_L$ partners is governed by three types of interactions: the weak interaction, the interaction with the Higgs boson and the self-interaction. Since the coupling constant of the weak interaction is already fixed by experiments performed so far, there are two undetermined couplings, $\lambda_{\chi H}$ and $\lambda_\chi$, in addition to an undetermined mass parameter $m_\chi$ in the theory. Since phenomenology of the dark matter is governed by $m_\chi$ and $\lambda_{\chi H}$, while not relevant to $\lambda_\chi$ at present, we consider the range of the first two parameters. Moreover, we focus on the range so that $\lambda_{\chi H}$ is positive to make the discussion of the vacuum stability simple.\footnote{Since $\lambda_{\chi H}$ and $\lambda_\chi$ are always positive up to high enough scale when $\lambda_{\chi H} >0$ and $m_\chi$ is in the ${\cal O}(1)$\,TeV scale, our vacuum becomes enough stable compared to the age of the universe\,\cite{Buttazzo:2013uya}. In passing, the coupling constant of the weak interaction does not diverge up to the high-energy scale in the scalar triplet theory.} 

\begin{figure}[t]
    \centering
    \includegraphics[keepaspectratio, scale=0.62]{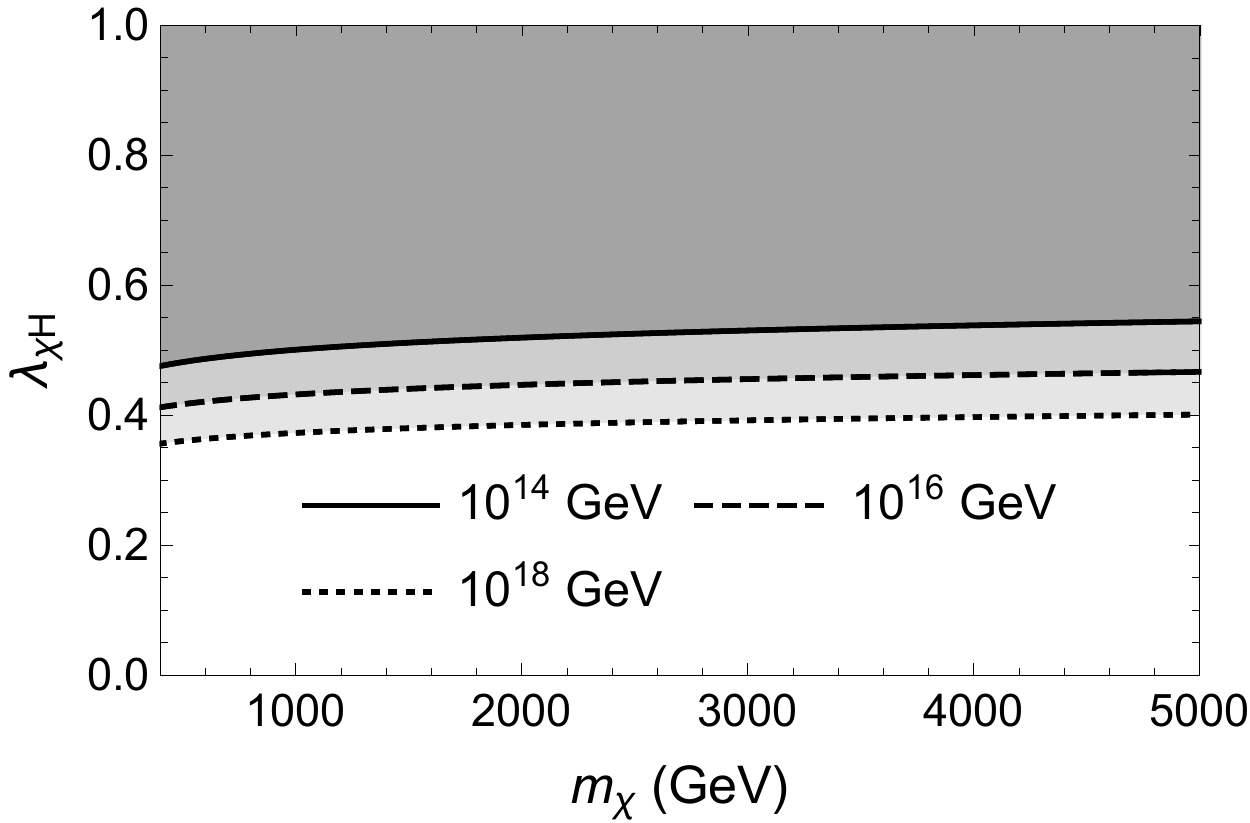}
    \caption{\small \sl The parameter region that the scalar triplet theory does not break down up to the energy scale of $10^{18}$, $10^{16}$ and $10^{14}\,{\rm GeV}$ are depicted as white, light grey and gray regions, respectively.}
    \label{LPfig}
\end{figure}

When we postulate that the theory of the scalar triplet dark matter does not break down up to high energy scale, the parameters are required to be in a certain region. By integrating RGEs of the theory with two-loop  $\beta$ functions shown in Appendix\,\ref{appendix: beta function}, such a region is obtained as depicted in Fig.\,\ref{LPfig}, where the coupling constant $\lambda_{\chi H}$ is defined at the scale of $m_\chi$. In order to figure out the region, we match two types of RGEs at the scale $m_\chi$: RGEs of the SM with appropriate initial conditions of the SM couplings below the scale\,\cite{Buttazzo:2013uya}, while those including the contribution of the scalar triplet with the initial condition $\lambda_\chi (m_\chi) = 0$ above the scale. The initial condition of $\lambda_\chi$ is optimized in order to make the preferred parameter region conservative\,\cite{Hamada:2015bra}, namely it makes the preferred parameter region of $\lambda_{\chi H}$ largest at a given $m_\chi$. As seen in the figure, the theory does not break down if $\lambda_{\chi H}$ is less than 0.4--0.5.

\section{Thermal relic abundance}
\label{sec: abundance}

One of the important observables at WIMP dark matter scenarios is its thermal relic abundance. In this article, we focus on the WIMP dark matter scenario that does not lead to overclosure of the universe assuming the standard cosmological evolution, namely the relic abundance is equal to or less than the dark matter density observed today. We introduce (co-)annihilation processes that the triplet scalar dark matter theory predicts, discuss those annihilation cross sections including the Sommerfeld effect, and figure out the model parameter region that is attractive from the viewpoint of the thermal relic abundance.

\subsection{(Co-)annihilation processes}
\label{subsec: Annihilation}

Since all components in the multiplet $\chi$ are highly degenerate, not only the self-annihilation of the dark matter component $\chi^0$ but also coannihilation among $\chi^0$ and $\chi^\pm$ play important roles to determine the thermal relic abundance. The most important quantity is thus the thermally averaged effective annihilation cross section $\langle \sigma_{\rm eff} v \rangle$\,\cite{Griest:1990kh}, which is defined as
\begin{align}
    \label{thermallyaveraged}
    \langle \sigma_{\rm eff}\,v \rangle & \equiv \sum_{i,j} \frac{\langle \sigma_{ij} v \rangle}{g_{\rm eff}^2(x)}
    (1 + \Delta_i)^{3/2} (1 + \Delta_j)^{3/2} \exp \left[ - x\,(\Delta_i + \Delta_j ) \right], \nonumber \\
    g_{\rm eff}(x) & \equiv 1 + 2\prn{1 + \Delta_{\chi^\pm}}^{3/2} \exp \left[ - x\, \Delta_{\chi^\pm} \right],
\end{align}
where $i$ and $j$ run over $\chi^0$ and $\chi^\pm$, while $\Delta_{\chi^0} = 0$ and $\Delta_{\chi^\pm} = \delta m/m_{\chi^0}$ with $\delta m$ being the one in eq.\,(\ref{eq: mass difference}). The annihilation cross section between $\chi_i$ and $\chi_j$ is denoted by $\sigma_{ij}\,(= \sigma_{ji})$.

We now consider $\sigma_{ij}$ of each (co-)annihilation process at tree-level. Since the dark matter and its SU(2)$_L$ partners are scalar particles, all (co-)annihilation processes into a pair of SM fermions are p-wave suppressed and dominant contributions are from those into SM bosons. Moreover, since $\chi^0$ and $\chi^\pm$ are expected to be much heavier than the electroweak scale as we will see in the following subsections, the processes utilizing vertices $\chi \chi' h$\,(with $\chi$, $\chi'$ being $\chi^0$, $\chi^\pm$) are suppressed by $m_W/m_{\chi^0}$. Hence, dominant (co-)annihilation processes are those into $W^- W^+$, $h h$ for $\chi^0 \chi^0$, into $W^\pm Z$, $W^\pm A$ for $\chi^0 \chi^\pm$, into $W^\pm W^\pm$ for $\chi^\pm \chi^\pm$ and into $W^- W^+$, $Z Z$, $Z A$, $A A$, $h h$ for $\chi^+ \chi^-$,  with $A$, $h$, $W^\pm$ and $Z$ being photon, Higgs, $W$ and $Z$ bosons, respectively. The cross sections of the processes are obtained as follows:
\begin{align}
    \label{cstree}
    \sigma_{\chi^0 \chi^0} v &= 4 \pi \alpha_2^2/m_\chi^2 + \lambda_{\chi H}^2/(4 \pi m_\chi^2),
    \nonumber \\
    \sigma_{\chi^\pm \chi^\pm} v &= 2\,\sigma_{\chi^0 \chi^\pm} v = 2\pi \alpha_2^2/m_\chi^2,
    \nonumber \\
    \sigma_{\chi^+ \chi^-} v &= 3 \pi \alpha_2^2/m_\chi^2 + \lambda_{\chi H}^2/(4 \pi m_\chi^2),
\end{align}
with $\alpha_2 \equiv g_2^2/(4 \pi)$. To obtain the above cross sections, we neglect the effect of the mass difference $\delta m$ as well as the masses of the SM bosons at final states of the processes.

\subsection{Sommerfeld effect}
\label{subsec: Sommerfeld effect}

Since the dark matter under consideration has a weak charge, the (co-)annihilation processes discussed in the previous subsection are affected by the so-called the Sommerfeld effect\,\cite{Hisano:2004ds} when its mass is ${\cal O}(1)$\,TeV. The effect is intuitively understood as follows: When the dark matter is much heavier than the SM bosons and becomes non-relativistic (NR), the exchange of the bosons between dark matters (and/or SU(2)$_L$ partners) induces long-range forces. The forces then distort the wave functions of incident particles from the plane wave before annihilation, and significantly alter corresponding cross sections depending on the nature of the forces. From the field theoretical viewpoint, it is described as a non-perturbative effect summing all ladder diagrams exchanging the bosons at the NR limit.

The Sommerfeld effect is evaluated by solving appropriate Shr\"{o}dinger equations\,\cite{Hisano:2004ds, Cirelli:2007xd}, which is obtained utilizing the so-called NR lagrangian derived from the original lagrangian in eq.\,(\ref{lagrangian}). To evaluate the effect on the (co-)annihilation processes, we solve six Shr\"{o}dinger equations, namely those for states $\chi^+\chi^+$, $\chi^-\chi^-$, $\chi^0\chi^+$, $\chi^0\chi^-$, $\chi^0\chi^0$ and $\chi^+\chi^-$:
\begin{equation}
    \left[-\frac{\nabla_r^2}{m_\chi} + V_a(r) \right] \psi_a(r) = \frac{m_{\chi} v^2}{4} \psi_a(r),
\end{equation}
with $v$ being the relative velocity between the incident particles. Here, the index ‘a' distinguishes the states. The potential $V(r)$ for each Shr\"{o}dinger equation is given as follows:
\begin{align}
    & V_{++}(r) = V_{-\,-}(r) =  \frac{\alpha}{r} + c_W^2\,\alpha_2\frac{e^{-m_Z r}}{r} - \frac{\lambda_{\chi H}^2}{4 \pi}\frac{v_H^2}{m_\chi^2} \frac{e^{-m_h r}}{r} ,
    \nonumber \\
    & V_{0+}(r) = V_{0-}(r) = \alpha_2 \frac{e^{-m_w r}}{r} - \frac{\lambda_{\chi H}^2}{4 \pi} \frac{v_H^2}{m_\chi^2} \frac{e^{-m_h r}}{r} ,
    \nonumber \\
    & V_{00}(r) =
    \begin{pmatrix}
        0 &
        - \displaystyle \sqrt{2}\,\alpha_2 \frac{e^{-m_W r}}{r} \\
        - \displaystyle \sqrt{2}\,\alpha_2 \frac{e^{-m_W r}}{r} &
        - \displaystyle \frac{\alpha}{r} - c_W^2\,\alpha_2 \frac{e^{-m_Z r}}{r} + 2\delta m
    \end{pmatrix} 
    - \frac{\lambda_{\chi H}^2}{4 \pi}\frac{v_H^2}{m_\chi^2} \frac{e^{-m_h r}}{r}
    \begin{pmatrix}
        1 & 0 \\
        0 & 1
    \end{pmatrix} ,
    \nonumber \\
    & V_{+-}(r) =
    \begin{pmatrix}
        - \displaystyle \frac{\alpha}{r} - c_W^2\,\alpha_2 \frac{e^{-m_Z r}}{r} &
        - \displaystyle \sqrt{2}\,\alpha_2 \frac{e^{-m_W r}}{r} \\
        - \displaystyle \sqrt{2}\,\alpha_2 \frac{e^{-m_W r}}{r} & - 2\delta m
    \end{pmatrix} 
    - \frac{\lambda_{\chi H}^2}{4 \pi}\frac{v_H^2}{m_\chi^2} \frac{e^{-m_h r}}{r}
    \begin{pmatrix}
        1 & 0 \\
        0 & 1
    \end{pmatrix} ,
\end{align}
with $m_W$, $m_Z$ and $m_h$ being $W$, $Z$ and Higgs boson masses, respectively. Since the states $\chi^0\chi^0$ and $\chi^+\chi^-$ are mixed with each other, the potentials $V_{00}(r)$ and $V_{+-}(r)$ as well as corresponding wave functions $\psi_{00}(r)$ and $\psi_{+-}(r)$ are given by $2 \times 2$ matrices. Here, we omit tiny corrections arising from the mass difference $\delta m$ except those in the $2 \times 2$ potentials, for the latter could give a large impact on the Sommerfeld effect via the interference between $\chi^0\chi^0$ and $\chi^+\chi^-$. On the other hand, their boundary conditions are given as follows:
\begin{align}
    & \{\psi_a(0)\}_{ij} = \delta_{ij} ,
    \quad
    \{\psi_a'(\infty)\}_{ij} = i\,[ m_\chi^2 v^2/4 - m_\chi \{V_a(\infty)\}_{ii} + i0^+ ]^{1/2}\,\{\psi_a(\infty)\}_{ij} .
\end{align}
The indices $i$ and $j$ run over 1 and 2 for $a = \chi^0\chi^0$ and $\chi^+\chi^-$, while $i = j = 1$ for other processes. The prime in the second condition denotes the derivative with respective to $r$.

Taking the coefficient of the out-going wave of the wave function obtained by the solution of the above Shr\"{o}dinger equation, $\{A_a\}_{1j} \equiv \lim_{r \to \infty} [\{\psi_a(r)\}_{1j}/\exp (i m_\chi v r /2)]$, the cross section of each annihilation process including the Sommerfeld effect is obtained as
\begin{equation}
    \sigma_a v = c_a\,\left( A_a \cdot \Gamma_a \cdot A_a^\dag \right)_{11},
\end{equation}
where $c_a$ is 2 when the initial two-body state is composed of identical particles, otherwise it is 1. On the other hand, the annihilation coefficient (matrix) denoted by $\Gamma_a$ in the formula is obtained by the annihilation cross sections at tree-level discussed in section\,\ref{subsec: Annihilation},\footnote{The off-diagonal element of the annihilation matrix, $\Gamma_{ij}$, is obtained by calculating the imaginary part of one-loop diagrams describing the conversion from the state $j$ to $i$ utilizing the so-called Cutkosky rules\,\cite{Cutkosky:1960sp}.}
\begin{align}
    &\Gamma_{++} = \Gamma_{-\,-} = \frac{\pi \alpha_2^2}{m_\chi^2} ,
    \quad
    \Gamma_{0+} = \Gamma_{0-} = \frac{\pi \alpha_2^2}{m_\chi^2} ,
    \quad
    \{\Gamma_{+-}\}_{ij} = \{\Gamma_{00}\}_{(3-i)\,(3-j)} ,
    \nonumber \\
    &\Gamma_{00} = 
    \frac{\pi \alpha_2^2}{m_\chi^2} \begin{pmatrix} 2 & \sqrt{2} \\ \sqrt{2} & 3 \end{pmatrix}
    + \frac{\lambda_{\chi H}^2}{8 \pi m_\chi^2} \begin{pmatrix} 1 & \sqrt{2} \\ \sqrt{2} & 2 \end{pmatrix} .
\end{align} 

\begin{figure}[t]
    \centering
    \includegraphics[keepaspectratio, scale=0.61]{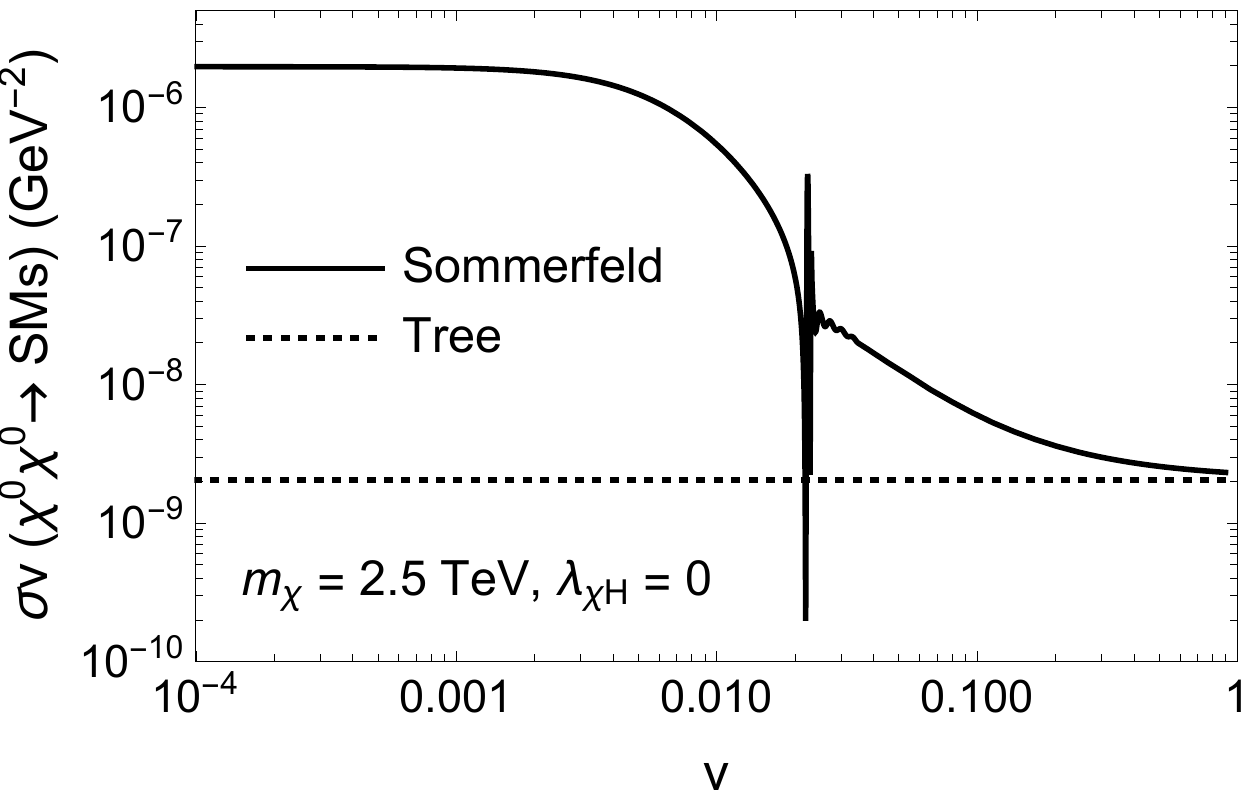}
    \quad
    \includegraphics[keepaspectratio, scale=0.61]{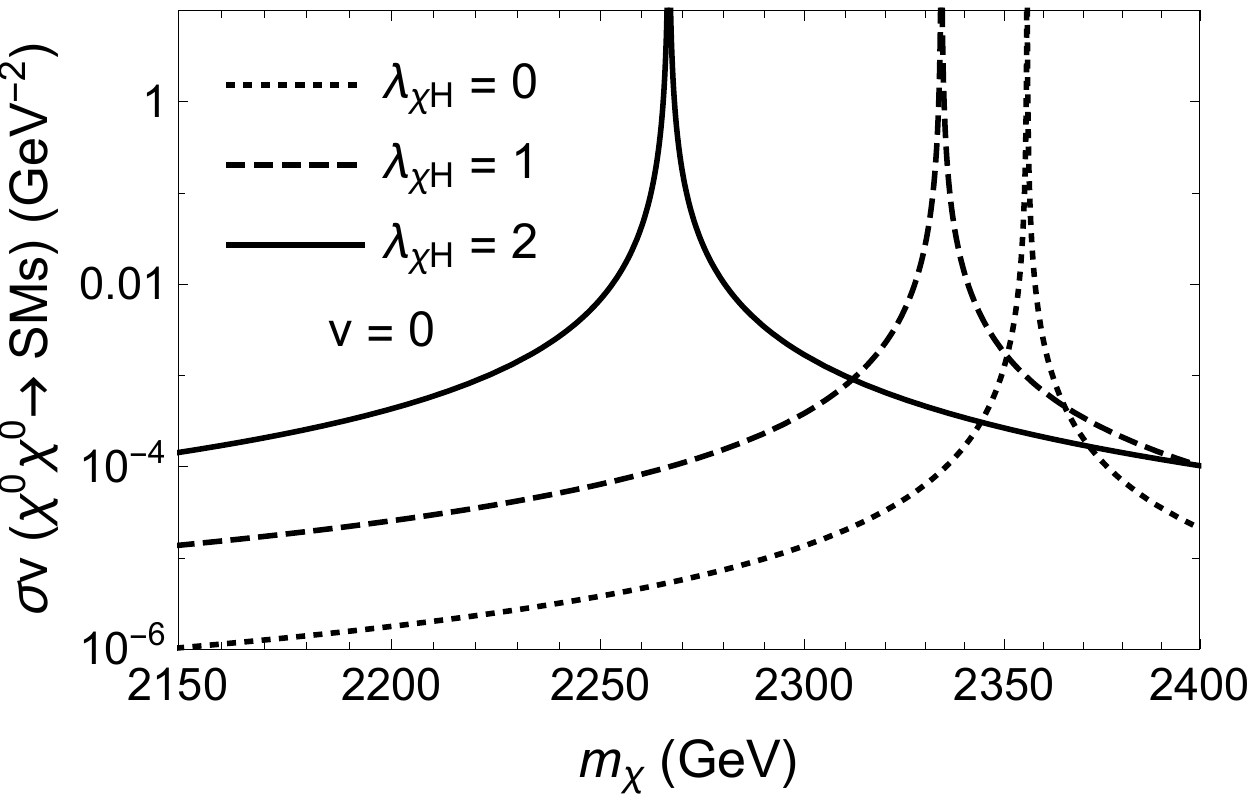} \\
    \vspace{0.5cm}
    \includegraphics[keepaspectratio, scale=0.61]{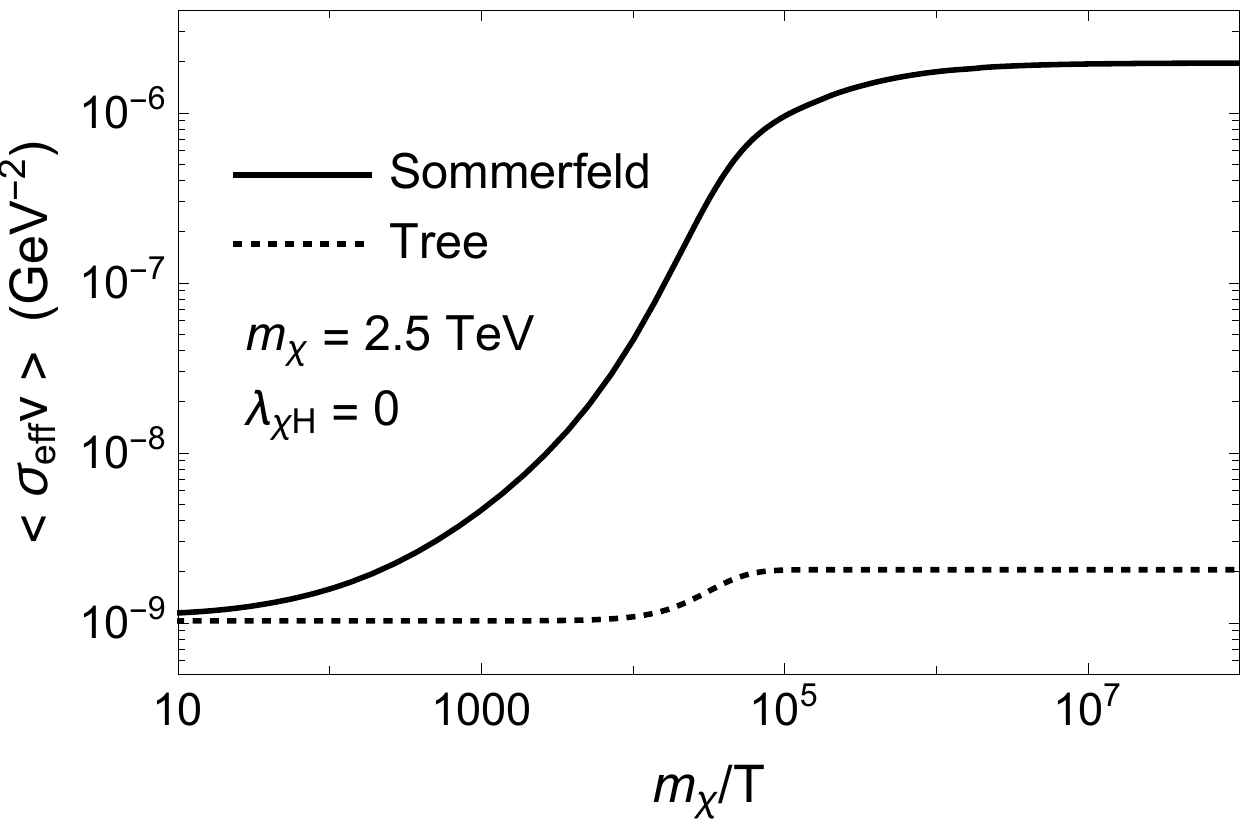}
    \quad
    \includegraphics[keepaspectratio, scale=0.61]{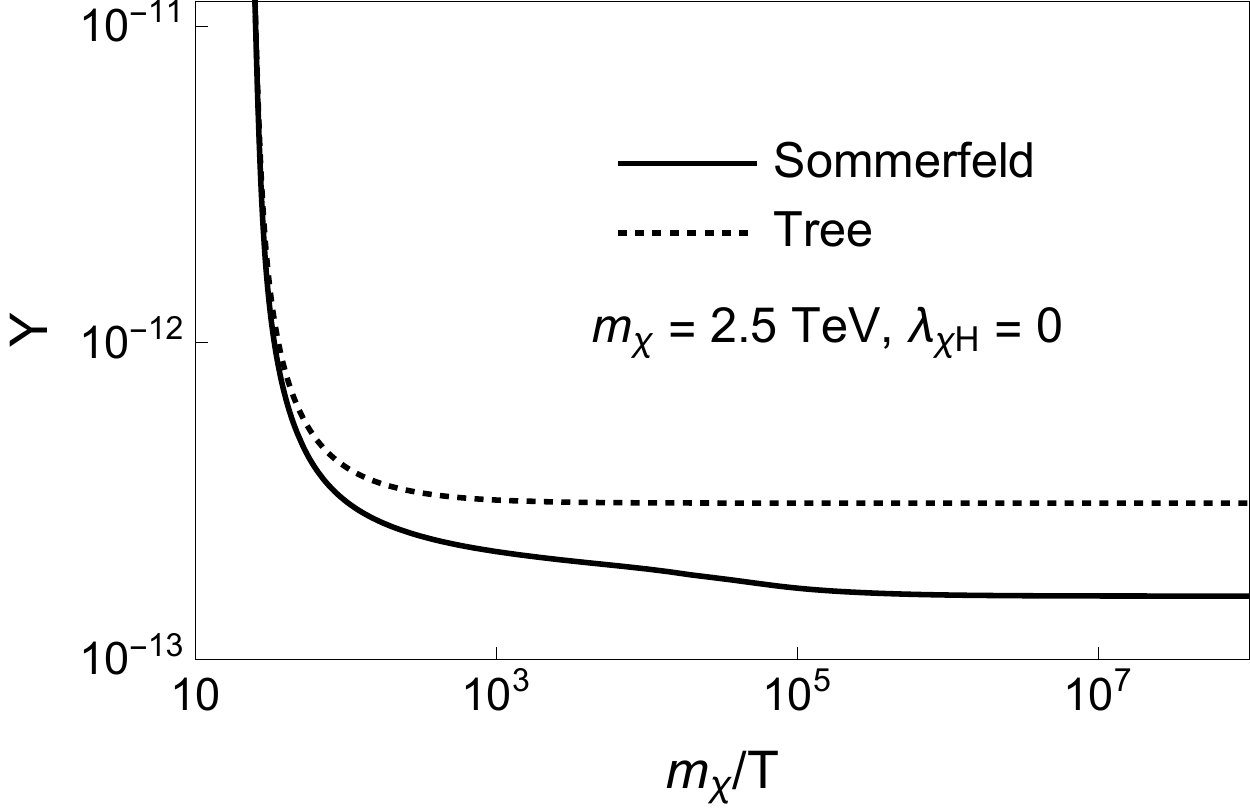}
    \caption{\small \sl The total annihilation cross section of the dark matter into SM particles as a function of the relative velocity with $m_\chi = 2.5$\,TeV \& $\lambda_{\chi H} = 0$ (top-left), the same cross section as a function of $m_\chi$ with $v = 0$ \& $\lambda_{\chi H} =$ 0, 1, 2 (top-right), the thermally averaged effective annihilation cross section as a function of the inverse temperature with $m_\chi = 2.5$\,TeV \& $\lambda_{\chi H} = 0$ (bottom-left), and the comoving density as a function of the temperature with $m_\chi = 2.5$\,TeV \& $\lambda_{\chi H} = 0$ (bottom-right).}
    \label{sigmafig}
\end{figure}

The total annihilation cross section of the dark matter into the SM particles, $\sigma v (\chi^0 \chi^0 \to {\rm SMs})$, is shown in the top-left panel of Fig.\,\ref{sigmafig} as a function of the relative velocity $v$ with $m_\chi$ and $\lambda_{\chi H}$ being 2.5\,TeV and 0, respectively.\footnote{We have used $\alpha_2$ at the scale of $m_Z$ and $m_\chi$ for the potential and annihilation coefficients, respectively.} It is seen from the figure that the Sommerfeld effect is larger in the lower velocity, while saturated at below $v \sim 0.01$. Because the kinetic energy of the two-body state $\chi^0\chi^0$, namely $m_{\chi}v^2/4$, cannot overcome the mass difference $2\,\delta m_\chi$ when $v \lesssim 0.01$, the state $\chi^0 \chi^0$ can not transfer into the other state $\chi^+ \chi^-$.

The same annihilation cross section but as a function of $m_\chi$ (instead, $v$ being fixed to be zero) is shown in the top-right panel of Fig.\,\ref{sigmafig} with several choices of $\lambda_{\chi H}$. The peak structure is seen at around $m_\chi \sim 2.3$\,TeV, and it indicates the existence of the zero-energy resonance at the mass. The position of the peak is intuitively understood as follows: When $\lambda_{\chi H} = 0$, it is determined by the condition that the rough estimate of the binding energy $\sim \alpha_2^2\,m_\chi$ is equal to the mass difference $2\delta  m_\chi \sim \alpha_2\,m_W$. It indeed leads to $m_\chi \sim$ 2.4\,TeV. When $\lambda_{\chi H} \neq 0$, the above binding energy is replaced by $\sim [\alpha_2 + \lambda_{\chi H}^2 v_H^2/(4 \pi m_\chi^2)]^2\,m_\chi$. Hence, the position of the peak is getting smaller when the larger $\lambda_{\chi H}$ is adopted, as can be seen in the figure.

The thermally averaged effective annihilation cross section defined in eq.\,(\ref{thermallyaveraged}) is shown in the bottom-left panel of Fig.\,\ref{sigmafig} as a function of the inverse temperature of the universe (normalized by the dark matter mass) $m_\chi/T$ with $m_\chi$ and $\lambda_{\chi H}$ being fixed to be 2.5\,TeV and 0, respectively. First, a slight increase of the tree level result
at $m_\chi/T \sim 10^5$ is due to the decoupling of the SU(2)$_L$ partners: The effective degrees of freedom $g_{\rm eff}$ is dropped from 3 to 1 (thus, $g_{\rm eff}^2$ drops from 9 to 1) as seen in eq.\,(\ref{thermallyaveraged}), while the sum of the cross sections $\sum_{ij} \sigma_{ij}$ drops from 18 to 4 as seen in eq.\,(\ref{cstree}), so that the effective annihilation cross section at the late time becomes twice larger than the one before the decoupling. Next, the effective annihilation cross section with the Sommerfeld effect keeps increasing as the temperature of the universe drops, and it is saturated at around $m_\chi/T \sim 10^5$ due to the decoupling of the SU(2)$_L$ partners; the typical velocity of the dark matter is well below 0.01 when $m_\chi/T > 10^5$, so that the Sommerfeld effect becomes constant, as discussed above.

\subsection{Constraint from the thermal relic abundance}
\label{Constraint from relic abundance}

We are now at the position to discuss constraints on the scalar triplet theory from its thermal relic abundance. Taking the temperature dependence of $\langle \sigma_{\rm eff}\,v \rangle$ into account, we integrate the following Boltzmann equation numerically according to the method in Ref.\,\cite{Griest:1990kh},
\begin{equation}
    \label{boltzmanneq}
    \frac{d Y}{d x} = -\frac{\langle \sigma_{\rm eff}\,v \rangle}{H\,x} \left( 1 - \frac{x}{3 g_{*s}}\frac{d g_{*s}}{d x} \right)\,s\,(Y^2 -Y_{\rm eq}^2),
\end{equation}
where $x \equiv m_\chi/T$, and $Y \equiv n/s$ is the number density per the comoving volume with $n$ and $s$ being the total number density of $\chi^0$ \& $\chi^\pm$ and the total entropy density of the universe, respectively. The Hubble parameter is denoted by $H$, while $Y_{\rm eq}$ is the value that $Y$ is expected to have when $\chi^0$ and $\chi^\pm$ are in the equilibrium. Relativistic degrees of freedom of the thermal bath, $g_{*}$ (used to calculate $H$ via the Friedman equation) and $g_{*s}$, are given as functions of temperature $T$ according to Ref.\,\cite{Saikawa:2018rcs,*Saikawa:2020swg}. With the present value of the comoving density $Y_0$, the relic abundance of the dark matter is obtained as $\Omega_{\rm DM}\,h^2 = m_\chi s_0 Y_0 h^2/\rho_c$ with $s_0$, $h \simeq 0.73$ and $\rho_c = 1.05 \times 10^{-5}\,h^2$\,GeV\,cm$^{-3}$ being the entropy density, the normalized Hubble constant and the critical density of the present universe, respectively\, \cite{Yao:2006px}.

The temperature dependence of the comoving density $Y(x)$ is shown in the bottom-right panel of Fig. \ref{sigmafig} with $m_\chi$ and $\lambda_{\chi H}$ being fixed to be 2.5\,TeV and 0, respectively. It is seen from the figure that the comoving density including the Sommerfeld effect keeps decreasing even after the freeze-out temperature, $T_f \sim m_\chi/20$, compared to that without the effect. Such a behavior is indeed expected from the temperature dependence of the thermally averaged effective annihilation cross section, $\langle \sigma_{\rm eff}\,v \rangle$, in the bottom-left panel of the figure, because it becomes larger when the temperature is lower as long as $x$ is smaller than $\sim 10^5$.

\begin{figure}[t]
    \centering
    \includegraphics[keepaspectratio, scale=0.61]{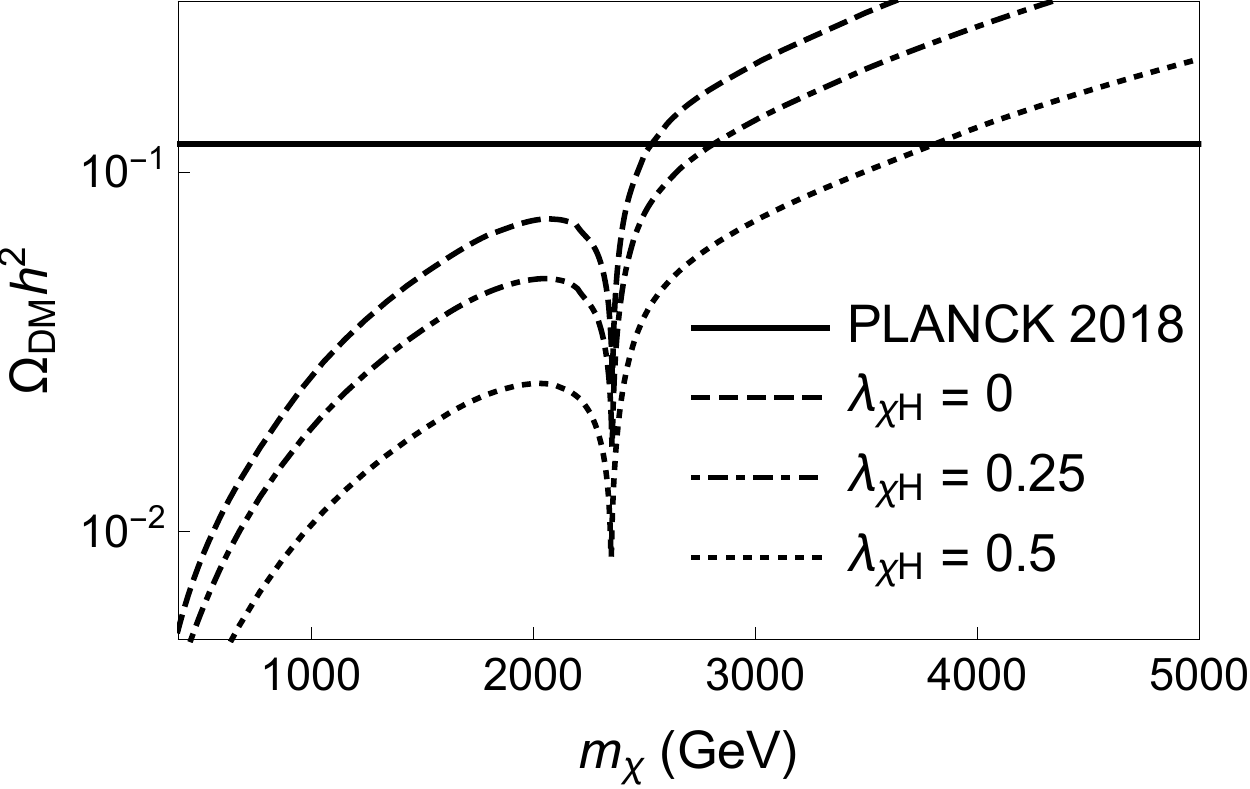}
    \quad
    \includegraphics[keepaspectratio, scale=0.61]{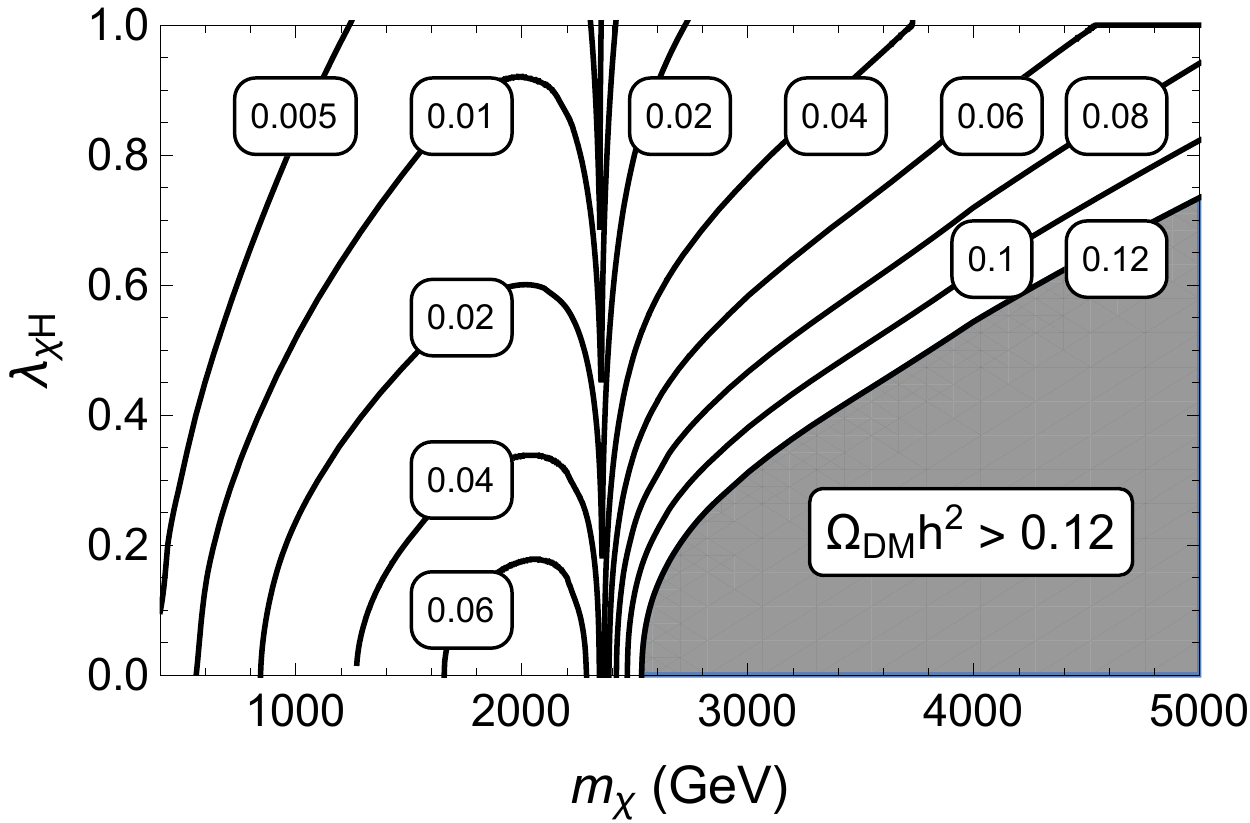}
    \caption{\small \sl {\bf (Left panel)} Thermal relic abundance of the dark matter $\Omega_{\rm DM}\,h^2$ as a function of $m_\chi$ for several choices of $\lambda_{\chi H}$. The horizontal line (PLANCK 2018) is the dark matter density observed today $\Omega_{\rm DM}^{\rm (obs)}\,h^2 \simeq 0.120$. {\bf (Right panel)} Contours of the abundance on the $(m_\chi, \lambda_{\chi H})$-plane. The region that the dark matter is overproduced, $\Omega_{\rm DM}\,h^2 \gtrsim 0.120$, is shown as a dark-gray shaded region.}
    \label{LPabundancefig}
\end{figure}

The thermal relic abundance of the dark matter $\Omega_{\rm DM}\,h^2$ is shown in the left panel of Fig.\,\ref{LPabundancefig} as a function of the dark matter mass for several choices of $\lambda_{\chi H}$, and it is compared with the dark matter density observed today $\Omega_{\rm DM}^{\rm (obs)}\,h^2 \simeq 0.120$\,\cite{Aghanim:2018eyx}. On the other hand, in the right panel of the figure, several contours of the abundance are depicted on the $(m_\chi, \lambda_{\chi H})$-plane. Assuming the standard cosmological evolution, the region that the scalar triplet dark matter is overproduced, i.e. $\Omega_{\rm DM}\,h^2 \geq \Omega_{\rm DM}^{\rm (obs)}\,h^2$, is given as a dark-gray shaded region. Both panels show a peculiar feature at $m_\chi \simeq 2.5$\,TeV due to the zero-energy resonance mentioned in the previous subsection. In the right panel, the observed dark matter abundance is fully explained by the thermal relics of the scalar dark matter when the parameters $m_\chi$ and $\lambda_{\chi H}$ are on the edge of the shaded region. On the other hand, the region above the edge means that the thermal relics are under-abundant to explain the dark matter density observed today, while it is still cosmologically viable if we consider e.g. the following scenarios:
\begin{itemize}
    \item[(I)] The scalar triplet dark matter contributes in part to the dark matter density observed today, and the rest of the density is from other dark matter candidates, e.g. Axion.
    \item[(II)] In addition to the thermal production process discussed above, a non-thermal process exists for the scalar triplet dark matter that does not spoil the standard cosmological evolution,\footnote{One of such processes is the decay of a very weakly interacting massive particle into the dark matter after the freeze-out. For instance, gravitino or moduli in supersymmetric scenarios can be such a massive particle.} and the whole density observed today is from of the scalar dark matter. 
\end{itemize}
We discuss direct and indirect detections of the real scalar triplet dark matter in following sections assuming the two cosmological scenarios mentioned above in order to quantitatively depict present constraints as well as projected sensitivities on the dark matter.

\section{Detection of the scalar triplet dark matter}  
\label{sec: detection}

We discuss here present constraints and future projected sensitivities on the search for the scalar triplet dark matter in various detection experiments and observations. For thermal dark matter candidates such as the one that we are discussing in this article, we usually think about three strategies for the detection; collider, direct and indirect detection.

In the detection of the scalar triplet dark matter at collider experiments, the disappearing charged track search is known to be the most sensitive one, where the track is from the charged SU(2)$_L$ partner $\chi^\pm$ decaying into the dark matter $\chi^0$ and a pion with the decay length of 6--7\,cm (due to the strong degeneracy in mass between $\chi^\pm$ and $\chi^0$). Then, the LHC experiment has excluded the dark matter lighter than $\sim$300\,GeV at the present 13\,TeV running, and it will be extended to $\sim$500\,GeV in the near future (HL-LHC) if no new physics signal is discovered\,\cite{Chiang:2020rcv}. On the other hand, since the most interesting parameter region of the dark matter mass is ${\cal O}(1)$\,TeV as seen in section\,\ref{sec: abundance} and it is much away from the projected sensitivity in the near future, we do not discuss the collider detection anymore. It is however worth pointing out that future collider experiments such as the 100\,TeV collider experiment have the projected sensitivity to search for the dark matter with ${\cal O}(1)$\,TeV mass\,\cite{Chiang:2020rcv}.

We therefore discuss the detection of the scalar triplet dark matter at underground experiments (i.e. the direct dark matter detection) and astrophysics observations (i.e. the indirect dark matter detection) in the following subsections in some detail, because those seem possible to search for the triplet dark matter with ${\cal O}(1)$\,TeV mass in the near future.

\subsection{Direct dark matter detection}

\begin{figure}[t]
    \centering
    \includegraphics[keepaspectratio, scale=0.5]{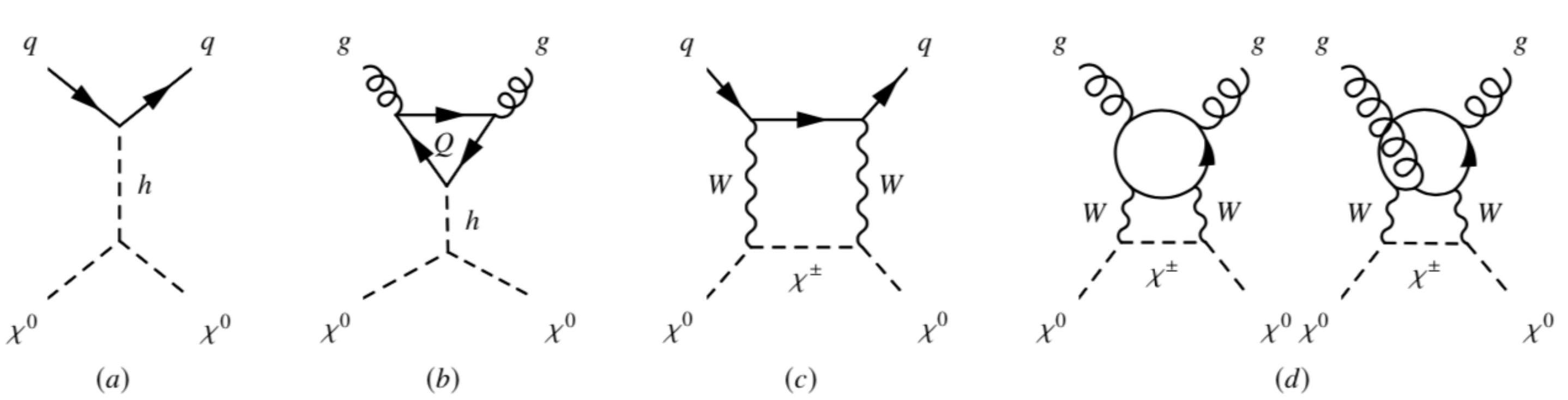}
    \caption{\small \sl Feynman diagrams inducing effective interactions between the WIMP and a quark/gluon.}
\label{diagramfig}
\end{figure}

Severe constraints on WIMP dark matter candidates are obtained in general by the direct dark matter detection searching for the scattering between the WIMP and a nucleus at underground laboratories. Given that the momentum transfer is enough low, the scattering is known to be well described by several low-energy effective interactions\,\cite{Hisano:2017jmz}: For the case of the scalar triplet model, diagrams shown in Fig.\,\ref{diagramfig} mainly contribute to the interactions. First two diagrams give a leading contribution to the scattering when the coupling $\lambda_{\chi H}$ is not suppressed, while others become important when this coupling is suppressed.

At the leading order, the diagram\,(a) in Fig.\,\ref{diagramfig} induces the effective quark scalar interaction $(\chi^0)^2m_q\bar{q}q$ with $q$ being a light quark (i.e. $u$, $d$ or $s$ quark), while the diagram\,(b) gives the gluon scalar interaction $(\chi^0)^2\,G^a_{\mu \nu} G^{a\,\mu\nu}$. On the other hand, at the next-leading order, the diagram\,(c) induces the quark twist-two interaction $\chi^0 (i\partial^\mu) (i\partial^\nu) \chi^0\,O_{\mu\nu}^q$.\footnote{Here, we have confirmed that other one-loop diagrams inducing interactions between the WIMP and a quark are suppressed by $m_W/m_\chi$ or simply contribute to the renormalization of $\lambda_{\chi H}$, so that we neglect those.} Finally, the diagrams\,(d) contribute to the aforementioned gluon scalar interaction and the gluon twist-two interaction. The coefficient of the latter interaction is, however, suppressed by $\alpha_s$ compared to the former one, so that we ignore its contribution. Then, the effective lagrangian describing interactions between the WIMP and a quark/gluon is obtained as\footnote{The diagrams\,(d) in Fig.\,\ref{diagramfig} can be efficiently evaluated using the so-called Fock-Schwinger gauge\,\cite{Hisano:2010ct}.}
\begin{align}
    &\mathcal{L}_{\rm eff} = \sum_{u, d, s}
    \left[
        \frac{\lambda_{\chi H}}{m_h^2} (\chi^0)^2 m_q \bar{q} q
        + \frac{\pi\alpha_2^2}{m_\chi m_W^3} \chi^0 (i\partial^\mu)(i\partial^\nu) \chi^0 O_{\mu\nu}^q
    \right]
    - \frac{\alpha_s}{4\pi} \left(\frac{\lambda_{\chi H}}{m_h^2} - f^g_2 \right) (\chi^0)^2 G^a_{\mu\nu}G^{a\mu\nu},
    \nonumber \\
    &O_{\mu\nu}^q \equiv \frac{1}{2} \bar{q} i \left(D_\mu \gamma_\nu + D_\nu \gamma_\mu - \frac{1}{2} g_{\mu\nu} \slashed{D} \right) q,
    \qquad
    f^g_2 = \frac{5\pi \alpha_2^2 m_\chi}{9 m_W^3}
    \left[ 1 + \frac{1}{2(1+m_t/m_w)^2} \right].
    \label{efflagqg}
\end{align}
We took the leading term in $m_W/m_\chi$ for the coefficient of the twist-two operator, while its full result can be found in Ref.\,\cite{Chao:2018xwz}. We also took the leading term in $m_W/m_\chi$ for the coefficient $f^g_2$. Then, the scattering cross section between $\chi^0$ and a nucleon $N$ is given,
\begin{equation}
    \sigma_{\chi N} = \frac{\mu_{\chi N}^2}{4\pi} \frac{m_N^2}{m_\chi^2}
    \left[ \frac{2\lambda_{\chi H}}{m_h^2} f^N -\frac{4}{9} f^g_2\,f_{TG}^N + \frac{3\pi \alpha_2^2}{4} \frac{m_\chi}{m_w^3}\,f_{\rm PDF}^N \right]^2
    \label{sigmasi}
\end{equation}
where $\mu_{\chi N} \equiv m_\chi m_N/(m_\chi + m_N)$ is the reduced mass, while $f^N = 0.284\,(0.287)$\,\cite{Belanger:2013oya}, $f_{TG}^N = 0.917\,(0.921)$ and $f_{\rm PDF}^N = 0.526\,(0.526)$\,\cite{Hisano:2015rsa} are hadron matrix elements that are required for the above scattering cross section when the nucleon $N$ is a neutron\,(proton).

\begin{figure}[t]
    \centering
    \includegraphics[keepaspectratio, scale=0.61]{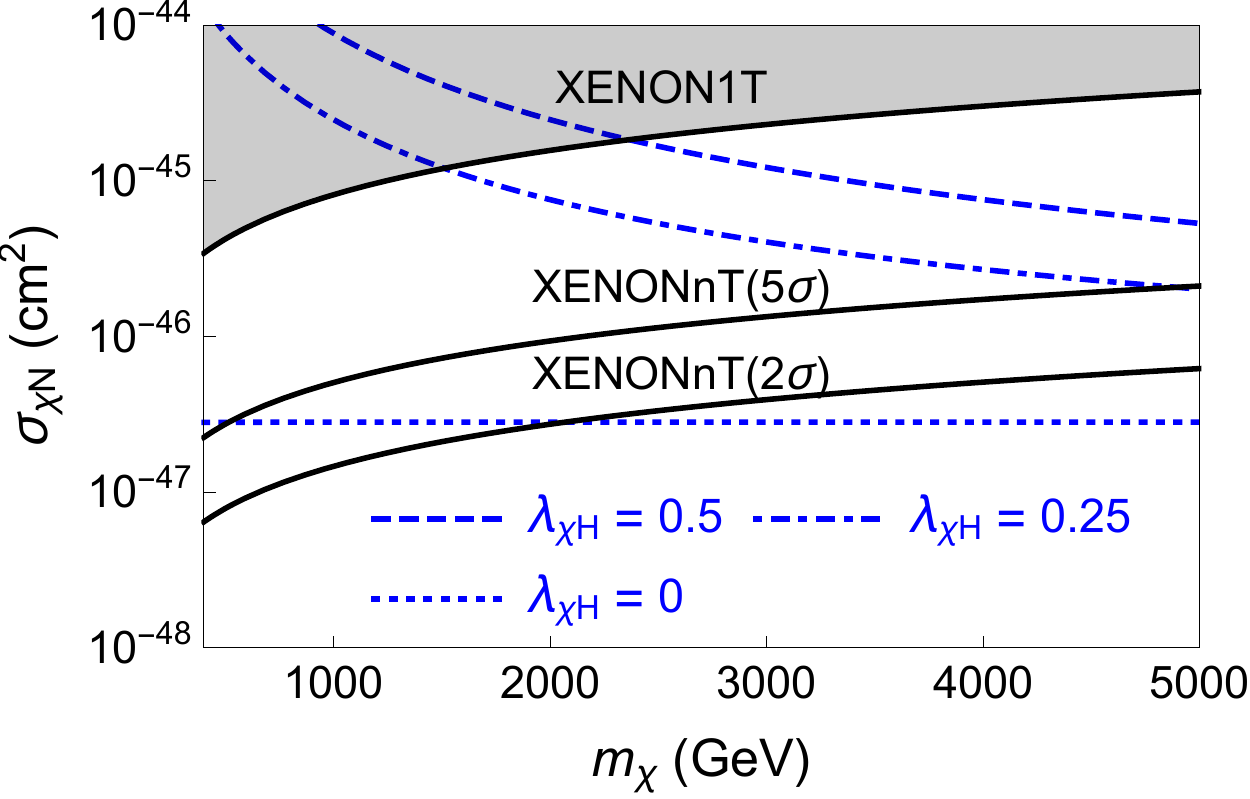}
    \quad
    \includegraphics[keepaspectratio, scale=0.61]{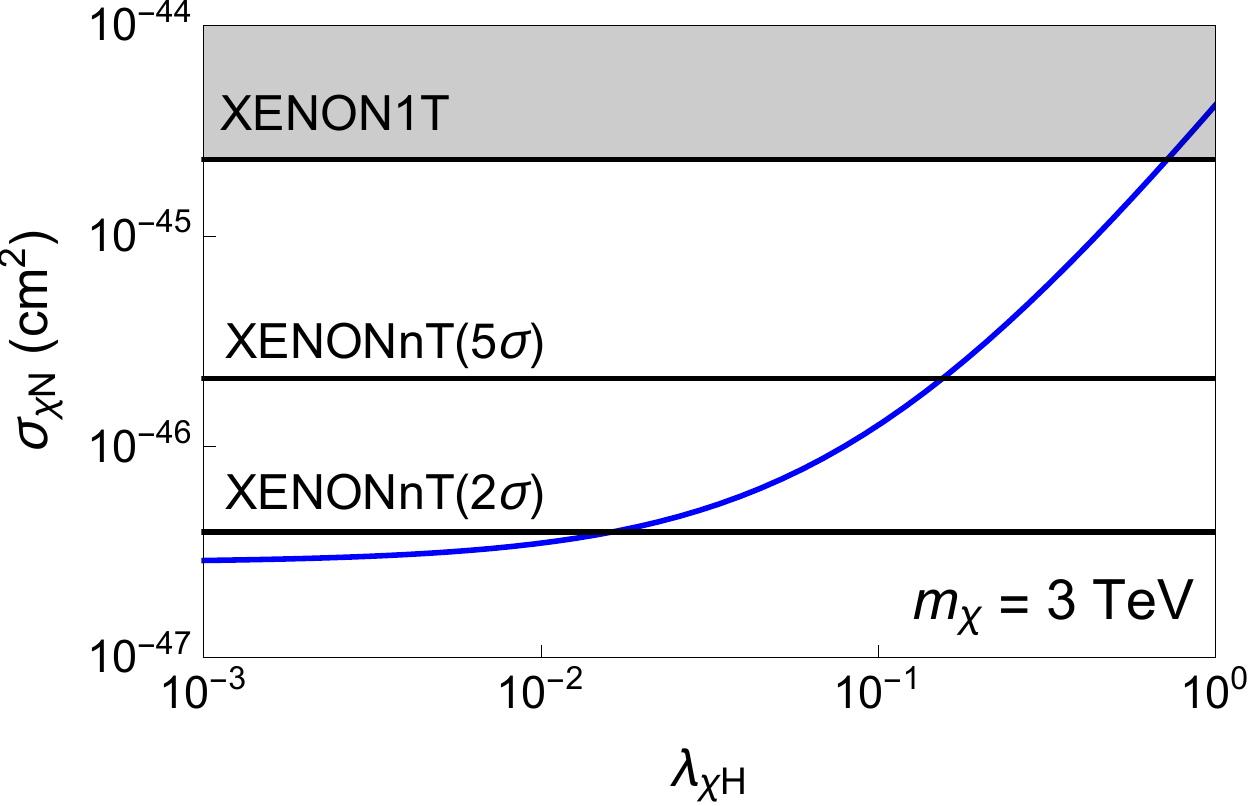}
    \caption{\small \sl The scattering cross section as a function of the dark matter mass $m_\chi$ for several choices of the coupling $\lambda_{\chi H}$ (left panel) and as a function of $\lambda_{\chi H}$ with $m_\chi$ being fixed to be 3\,TeV (right panel). The cross section is compared with the present constraint from the XENON1T experiment\,\cite{Aprile:2018dbl} and the future expected sensitivity of the XENONnT experiment at 2$\sigma$ and 5$\sigma$ levels\,\cite{Aprile:2020vtw} in both panels.}
    \label{fig: scattering}
\end{figure}

The scattering cross section is shown in Fig.\,\ref{fig: scattering} as a function of the dark matter mass $m_\chi$ for several choices of the coupling $\lambda_{\chi H}$ (left panel) and as a function of $\lambda_{\chi H}$ with $m_\chi$ being fixed to be 3\,TeV (right panel). The cross section is also compared with the present constraint from the XENON1T experiment\,\cite{Aprile:2018dbl} and the future expected sensitivity of the XENONnT experiment at 2$\sigma$ and 5$\sigma$ levels\,\cite{Aprile:2020vtw}. It is seen from the figure that the future experiment will search for the dark matter efficiently unless $\lambda_{\chi H}$ is highly suppressed. This fact means that, though last two terms in the parenthesis at the right-hand side of the equation\,(\ref{sigmasi}) are not numerically large, it becomes important to estimate the sensitivity of future direct detection experiments on the scalar triplet dark matter, in particular, when $\lambda_{\chi H}$ is suppressed.

\begin{figure}[t]
    \centering
    \includegraphics[keepaspectratio, scale=0.61]{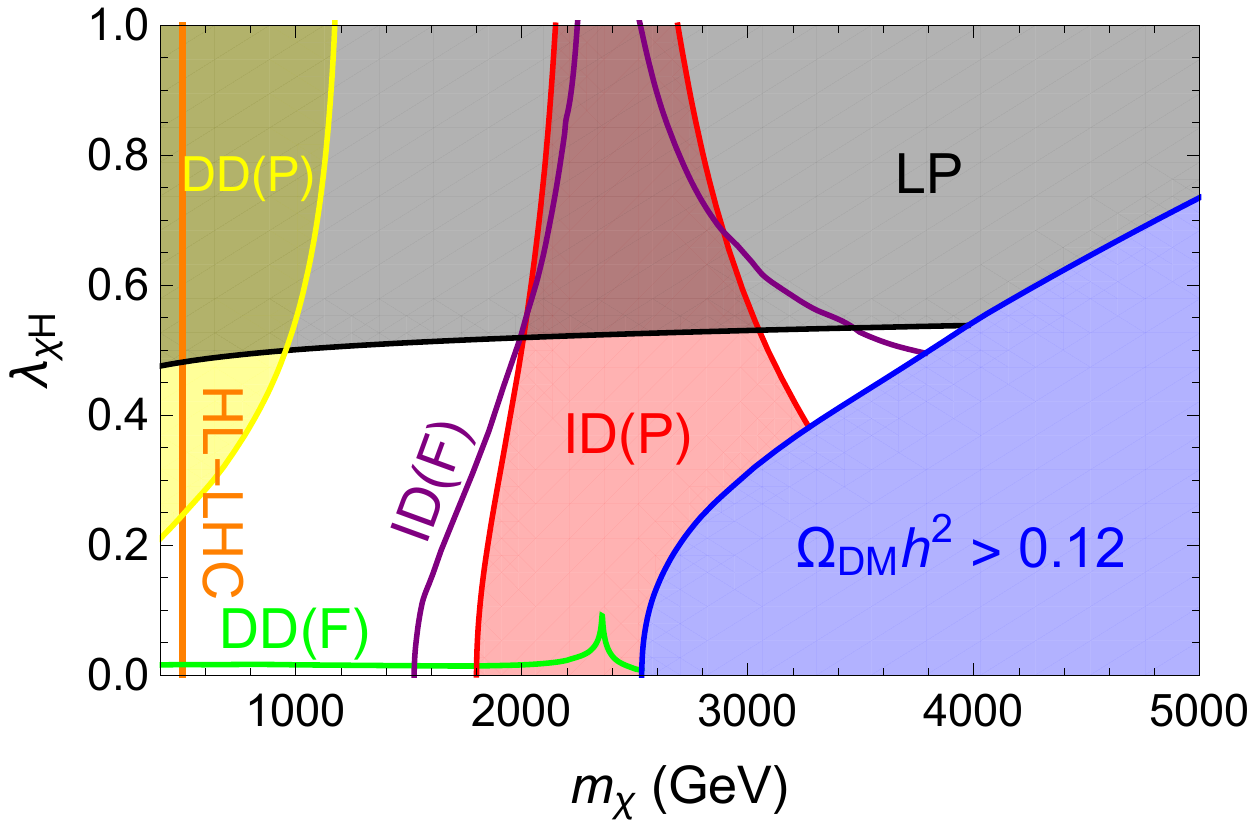}
    \quad
    \includegraphics[keepaspectratio, scale=0.61]{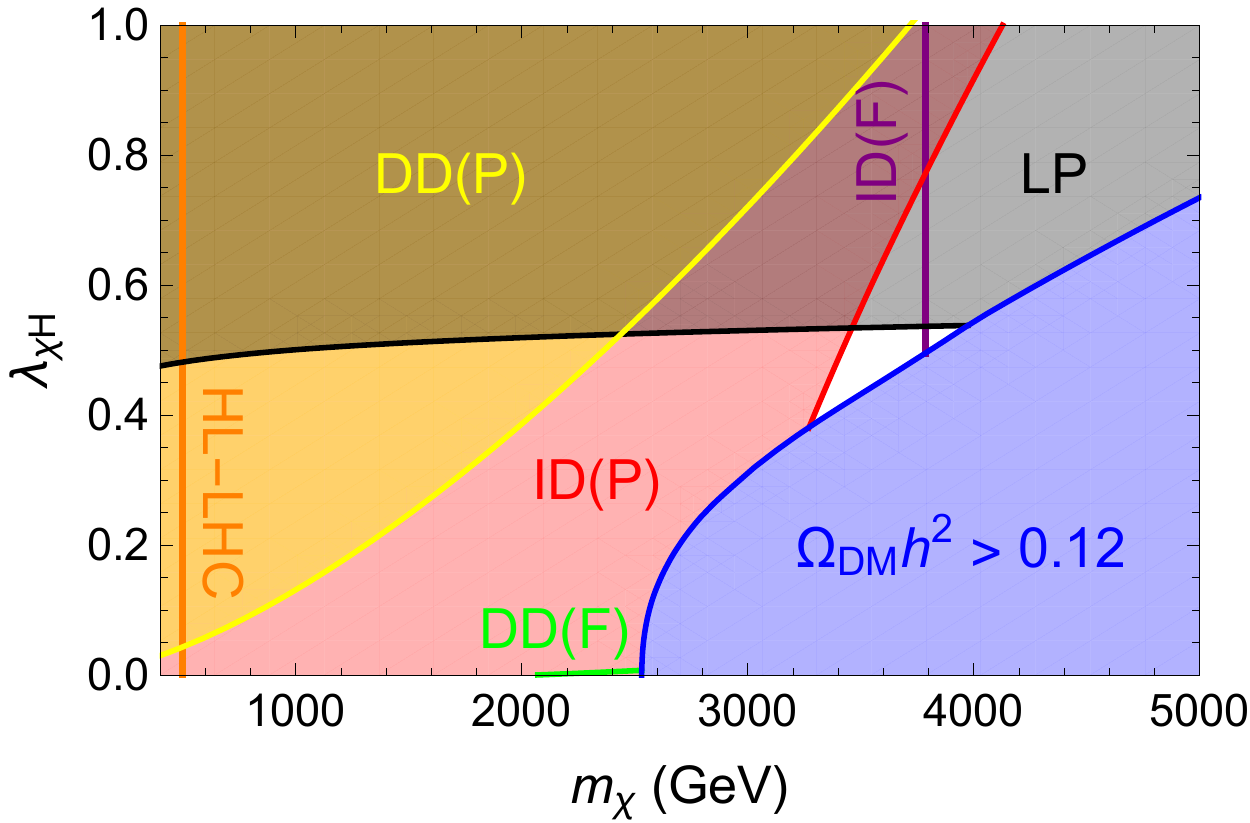}
    \caption{\small \sl Present constraints on the real scalar triplet dark matter from direct (XENON1T) and indirect (Fermi-LAT) detections are shown by yellow-shaded (named `DD(P)’) and red-shaded (named `ID(P)’) regions, respectively. The left and right panel are for the cosmological scenario (I) and (II) that are defined in section\,\ref{Constraint from relic abundance}. Future expected sensitivity to search for the dark matter at collider (HL-LHC), direct (XENONnT) and indirect (CTA) detections are shown by orange (named `HL-LHC'), green (named `DD(F)') and purple (named `ID(F)') lines, respectively. The region that the thermal relic abundance of the dark matter exceeds the dark matter density observed today, i.e. $\Omega_{\rm DM} h^2 > 0.1$, is shown by the blue painted region. On the other hand, the region that the theory breaks down below the high energy scale ($10^{14}$\,GeV) is shown by the black shaded region (named `LP’).}
    \label{constraintfig}
\end{figure}

Present constraint on the scattering cross section from the XENON1T experiment\,\cite{Aprile:2018dbl} gives a restriction on the model parameters as shown by a yellow-shaded region in Fig.\,\ref{constraintfig} named `DD\,(P)'. Here, the left panel is given for the cosmological scenario (I) defined in section\,\ref{Constraint from relic abundance}; the scalar triplet dark matter contributes in part to the whole dark matter density observed today according to the thermal production, while the right panel is for the scenario (II); the whole dark matter  density observed today is explained by the scalar triplet dark matter assuming the existence of a non-thermal production in addition to the thermal one. In the former case, the constraint from the direct detection experiment is applied not to the scattering cross section $\sigma_{\chi N}$ but to the scaled scattering cross section $(\Omega_{\rm DM}/\Omega_{\rm obs})\,\sigma_{\chi N}$ with $\Omega_{\rm obs}$ and $\Omega_{\rm DM}$ being the dark matter abundance observed today and the abundance of the scalar triplet dark matter from the thermal production process discussed in section\,\ref{Constraint from relic abundance}.

The expected sensitivity of the XENONnT experiment\,\cite{Aprile:2020vtw}, which is defined as the expected constraint on the parameter region at $2\,\sigma$ level assuming no signal is detected there, is also shown in both panels of Fig.\,\ref{constraintfig} as a green line named `DD\,(F)'. Here, the region above the line is covered by this near future experiment. Focusing on the preferred parameter area that the theory does not break down up to high energy scale (i.e. outside the black shaded region), the almost entire parameter region is proved by the experiment for both the cosmological scenarios (I) and (II), except tiny regions with very much suppressed $\lambda_{\chi H}$.

\subsection{Indirect dark matter detection}  
\label{sec: indirect}

WIMP dark matter candidates are also being searched for at the indirect dark matter detection observing energetic particles produced by their annihilation in the universe. Among various indirect detection methods, the one utilizing the gamma-ray observation of dwarf spheroidal galaxies (dSphs) enables us to search for the scalar triplet dark matter robustly and efficiently\,\cite{Lefranc:2016fgn, Arakawa:2021vih}, as the dSphs are dark matter rich objects having large $J$-factors (see below) that are estimated less ambiguously than those for other objects. Dominant processes of the dark matter producing such gamma-rays are $\chi^0 \chi^0 \to W^+ W^-$, $Z Z$ and $h h$ when the mass of the dark matter is enough heavier than the electroweak scale, which are followed by subsequent decays of $W^\pm$, $Z$ and $h$ bosons into photons with various energies.

The gamma-ray flux from the dark matter annihilation in a dSph is estimated to be
\begin{align}
    \frac{d\Phi_\gamma}{dE_\gamma} \simeq
    \left[
        \frac{\langle \sigma v \rangle_{\rm tot}}{8 \pi m_\chi^2} \sum_{f = W^+W^-,\,ZZ,\,hh}\,{\rm Br}\,(\chi^0 \chi^0 \to f)\,\left.\frac{d N_\gamma}{dE_\gamma}\right|_f
    \right]
    \times
    \left[
        \int_{\Delta \Omega} d\Omega \int_{\rm l.o.s} ds\,\rho_{\rm DM}^2
    \right],
    \label{flux}
\end{align}
where $\langle \sigma v \rangle_{\rm tot}$ is the velocity average of the total annihilation cross section (times the relative velocity) of the dark matter, ${\rm Br}\,(\chi^0 \chi^0 \to f)$ is the branching fraction of the annihilation into the final state `f' and $d N_\gamma/dE_\gamma|_f$ is the so-called fragmentation function describing the number of produced photons with energy $E_\gamma$ at a given final state `f'\,\cite{Cirelli:2010xx}. The cross section and the branching fractions are calculated in the same way as those in section\,\ref{subsec: Sommerfeld effect}. The term in the second parenthesis at the right-hand side of equation\,(\ref{flux}) is called the J-factor (of a dSph), which is, roughly speaking, the average of the dark matter mass density squared over the dSph. The uncertainty of the $J$-factor estimate is fortunately smaller than those of other targets\,\cite{Strigari:2013iaa} such as the $J$-factor of the center of Milky Way galaxy. We use the $J$-factors of dSphs adopted in Ref.\,\cite{Ackermann:2013yva} in our analysis, where the so-called Navarro-Frenk-White\,(NFW) profile\,\cite{Navarro:1996gj} is assumed as the distribution function of dark matters inside the dSphs.\footnote{Errors of the $J$-factors could be larger than those in the reference when we take into account some systematic uncertainties that are not considered for some dSphs\,(i.e. the non-sphericity of stellar and dark matter profiles\,\cite{Hayashi:2016kcy}, foreground contamination\,\cite{Ichikawa:2016nbi, Ichikawa:2017rph, Horigome:2020kyj}, spatial dependence of the velocity dispersion\,\cite{Ullio:2016kvy}, etc.). Hence, the real constraint from the indirect detection could be weaker than that given in this article.}

At present, the observation of the dSphs by the Fermi-LAT collaboration gives the most stringent constraint on the scalar triplet dark matter, where the low-energy tail of the continuum gamma-ray produced by the aforementioned annihilation processes are being searched for. The constraint on the model parameters is shown in Fig.\,\ref{constraintfig} as a red-shaded region named `ID\,(P)'. Here, to depict the region, we have used the constraint on the dark matter annihilation into $W^+W^-$ given in Ref.\,\cite{Ackermann:2015zua}, as the fragmentation function of the channel $d N_\gamma/dE_\gamma|_{W^+W^-}$ is almost the same as those of the other channels into $ZZ$, $hh$, i.e. $d N_\gamma/dE_\gamma|_{ZZ,\,hh}$, in the $E_\gamma$ region of interest when the dark matter mass is ${\cal O}(1)$\,TeV\,\cite{Cirelli:2010xx}.\footnote{For the constraint on the scalar triplet dark matter obtained by the Fermi-LAT observation, photons whose energy of ${\cal O}$(10--100)\,GeV play an important role. Then, $hh$ and $ZZ$ annihilation channels provide a similar number of photons to that of the $WW$ channel in this energy range, when the dark matter mass is ${\cal O}(1)$\,TeV. Hence, the constraints on all the above modes are expected to be almost identical to each other\,\cite{Ambrogi:2018jqj}.}

For the the cosmological scenario (I) shown in the left panel of Fig.\,\ref{constraintfig}, where the Fermi-LAT constraint is applied to the scaled annihilation cross section $(\Omega_{\rm DM}/\Omega_{\rm obs})^2 \langle \sigma v \rangle_{\rm tot}$, the present indirect dark matter detection excludes the region with $m_\chi \sim$ 2--3\,TeV. This is because the Sommerfeld effect causes the so-called the zero-energy resonance at around the mass region while the amount (number density) of dark matters at present universe is not very much suppressed. Meanwhile, the lighter and heavier mass regions are not excluded because the amount of dark matters is suppressed due to the suppressed thermal relic abundance (i.e. due to the large annihilation cross section) and the large dark matter mass without the boosted annihilation cross section, respectively. On the other hand, for the cosmological scenario (II) shown in the right panel, where the constraint is applied to the non-scaled annihilation cross section $\langle \sigma v \rangle_{\rm tot}$, the detection excludes the almost all region with the dark matter mass below 3--4\,TeV, though it is still possible to find the surviving parameter region at $m_\chi \sim 3.5$\,TeV and $\lambda_{\chi H} \sim 0.4$, which involves the 'WIMP-Miracle' region that all the dark matter abundance observed today is entirely from the thermal relics.

In the near future, the CTA experiment\,\cite{Consortium:2010bc} will cover a broader parameter region of the scalar triplet dark matter. Since the experiment enables us to observe very energetic gamma-rays, i.e. up to ${\cal O}(10)$\,TeV, the monochromatic gamma-ray signal from the annihilation contributes more to the detection than that of the continuum gamma-ray\,\cite{Lefranc:2016fgn}. The annihilation cross section producing the line gamma-ray signal is obtained in the same way as those in section\,\ref{subsec: Sommerfeld effect} with corresponding annihilation coefficient $\Gamma_{\rm line}$ as follows:
\begin{align}
    \sigma_{\rm line} v = 2\left( A_{00}
    \cdot \Gamma_{\rm line} \cdot A_{00}^\dag \right)_{11},
    \qquad
    \Gamma_{\rm line} \simeq
    \Gamma_{\gamma\gamma} + \frac{1}{2}\Gamma_{\gamma Z} = \frac{\pi \alpha_2^2}{m_\chi^2} (s_w^4 + c_w^2 s_w^2)
    \begin{pmatrix}
        0 & 0 \\
        0 & 2
    \end{pmatrix}.
\end{align}
Here, $m_\chi \gg m_W$ is assumed. Note that there will be a tree-level contribution to the $\{2,2\}$ component of the annihilation coefficient, i.e. $\{\Gamma_{\rm line}\}_{22}$, from the process $\chi^+ \chi^- \to \gamma h$ originating in the Higgs portal interaction $\lambda_{\chi H} \chi^2 |H|^2$. This contribution is, however, suppressed by $m_W^2/m_\chi^2$ compared to the other contributions from the processes originating in the gauge interaction, i.e. $\chi^+ \chi^- \to \gamma \gamma$ and $\gamma Z$. Hence, we do not include it in our analysis.\footnote{There are also one-loop contributions to all the components of the annihilation coefficient $\Gamma_{\rm line}$. Since the amplitude $|\{A_{00}\}_{11}|$ is the same order as $|\{A_{00}\}_{12}|$ when $m_\chi \gg m_W$, the contributions are negligibly smaller than the tree-level ones at $\{\Gamma_{\rm line}\}_{22}$. Hence, we do not include the one-loop contributions in our analysis.}

Since the CTA experiment is utilizing air Cerenkov telescopes, it individually observes a few targets (dSphs) rather than observing the whole sky such as the Fermi-LAT experiment. Among various targets, the ultrafaint dSph named Segue\,1 is expected to be one of the ideal targets giving the largest $J$-factor\,\cite{Ando:2020yyk}. We hence consider the expected constraint on the scalar triplet dark matter, as future prospects of its indirect dark matter detection, assuming that no signal is detected by observing Segue\,1 at the CTA experiment with the exposure time of 500\,h\,\cite{Ando:2021jvn}.\footnote{The $J$-factor of Segue\,1 has been estimated in Ref.\,\cite{Ando:2021jvn} using the same (NFW) dark matter profile as but a different method from those adopted in the Fermi-LAT collaboration. In the latter case, the factor has been estimated to be $\log_{10} J\,[{\rm GeV}^2\,{\rm cm}^{-5}] = 19.5 \pm 0.29$ and taken into account as a prior probability function (with a Log-Gaussian form) in the analysis. On the other hand, in the former case, the factor has been estimated to be $\log_{10}J\,[{\rm GeV}^2\,{\rm cm}^{-5}] \simeq 19$, and this value has been directly used to estimate the dark matter signal.} The constraint on the model parameters at 2$\sigma$ level is shown in Fig.\,\ref{constraintfig} as a purple line named `ID\,(F)'. In the left panel for the cosmological scenario (I), it is seen that the excluded region is steadily extended from the present constraint if no dark matter signal is observed there. Importantly, the CTA experiment will cover a part of the `WIMP-Miracle' region at $m_\chi \sim$ 4\,TeV. This is also true for the cosmological scenario (II) shown in the right panel, i.e. the present constraint could be extended to $m_\chi \lesssim 4$\,TeV if no dark matter signal is detected there, and it includes a part of the `WIMP-Miracle' region.

\section{Summary and discussion}
\label{sec: conclusion}

We have studied an extension of the SM with a real scalar triplet on which a $\mathbb{Z}_{2}$ symmetry is imposed. The scalar triplet has two charged components $\chi^\pm$ and one neutral component $\chi^0$ which plays the role of dark matter. Here, the radiative correction automatically makes $\chi^\pm$ heavier than $\chi^0$ about 166\,MeV. Phenomenology of the theory is governed by two free parameters, the mass of dark matter $m_\chi$ and the coupling of the interaction between the dark matter and the Higgs boson $\lambda_{\chi H}$. By solving RGEs of interaction couplings predicted in the theory with two-loop $\beta$ functions, we found that $\lambda_{\chi H}$ should be smaller than $\sim 0.5$ when postulating that the theory does not break down up to enough high energy scale.

The result we have obtained is as follows: First, we have calculated the relic abundance of the scalar triplet dark matter including the Sommerfeld effect on the dark matter annihilation and evaluated constraints from direct and indirect dark matter searches at the underground (XENON1T) experiment and the astrophysical (Fermi-LAT) observations, respectively. We have found that a wide parameter region is still allowed even if the coupling $\lambda_{\chi H}$ is less than $\sim 0.5$, and interestingly it includes the so-called the `WIMP-Miracle' one.

\begin{figure}[t]
    \centering
    \includegraphics[keepaspectratio, scale=0.61]{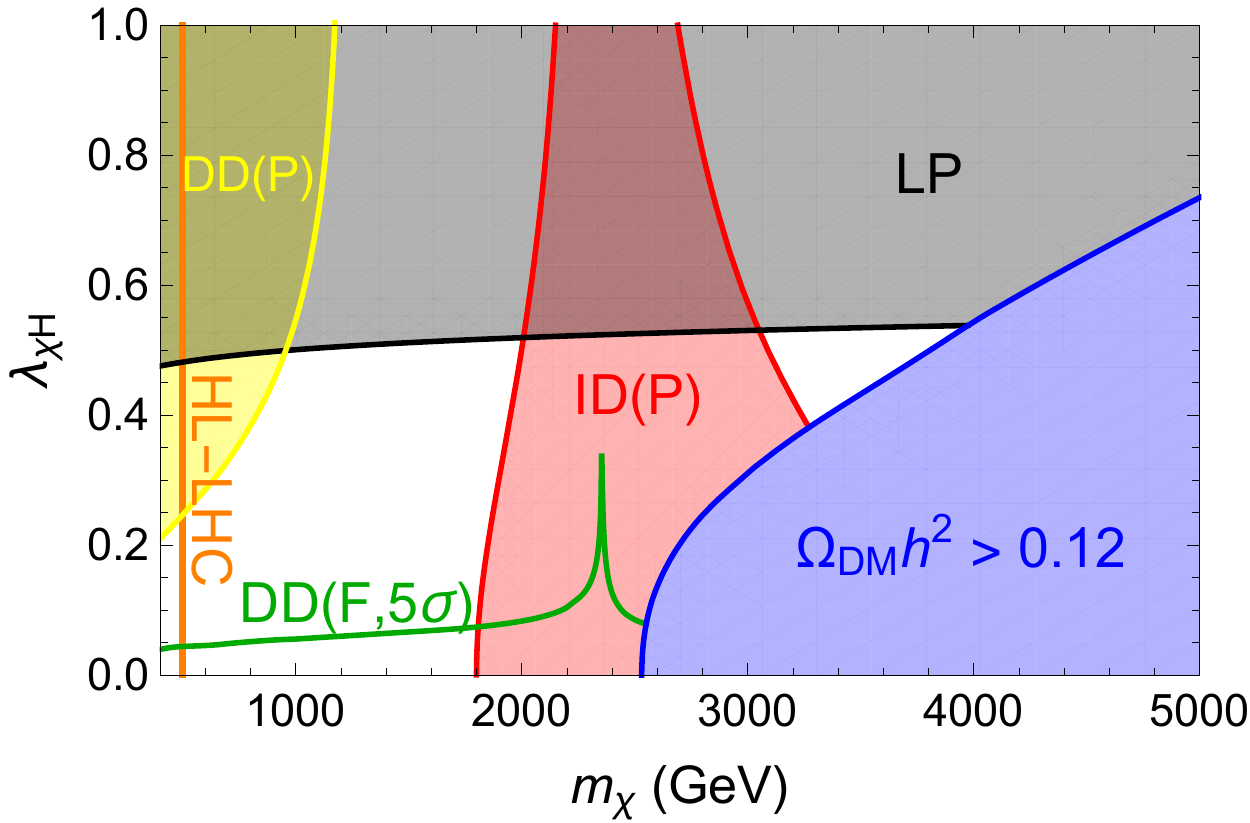}
    \quad
    \includegraphics[keepaspectratio, scale=0.61]{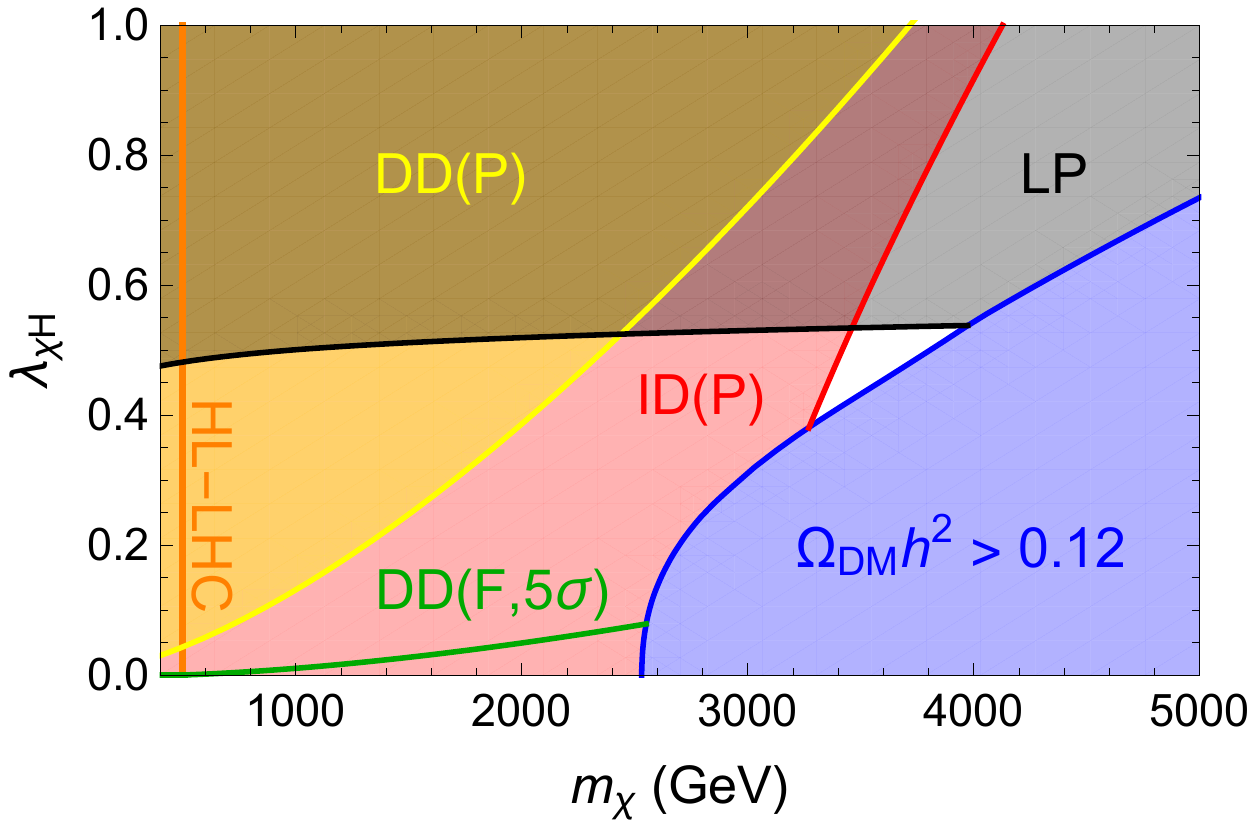}
    \caption{\small \sl The scalar triplet dark matter can be detected at 5$\sigma$ level in the region above the dark green line by XENONnT. All shaded regions and the orange line are the same as those in Fig.\,\ref{constraintfig}.}
    \label{dd5sigmafig}
\end{figure}

Next, with the fact that the scalar triplet dark matter lighter than 500\,GeV is explored at the HL-LHC experiment, the rest of the allowed parameter region will be efficiently searched for at near-future direct dark matter detection experiments. In particular, the XENONnT experiment will cover the almost entire parameter region as seen in Fig.\,\ref{constraintfig}. Moreover, this near future experiment will also allow us to discover the scalar triplet dark matter in a wide parameter region including the `WIMP-Miracle' one as shown by a dark green line in Fig.\,\ref{dd5sigmafig}, where the dark matter can be detected at the level of 5$\sigma$ in the region above the line.

\begin{figure}[t]
    \centering
    \includegraphics[keepaspectratio, scale=0.61]{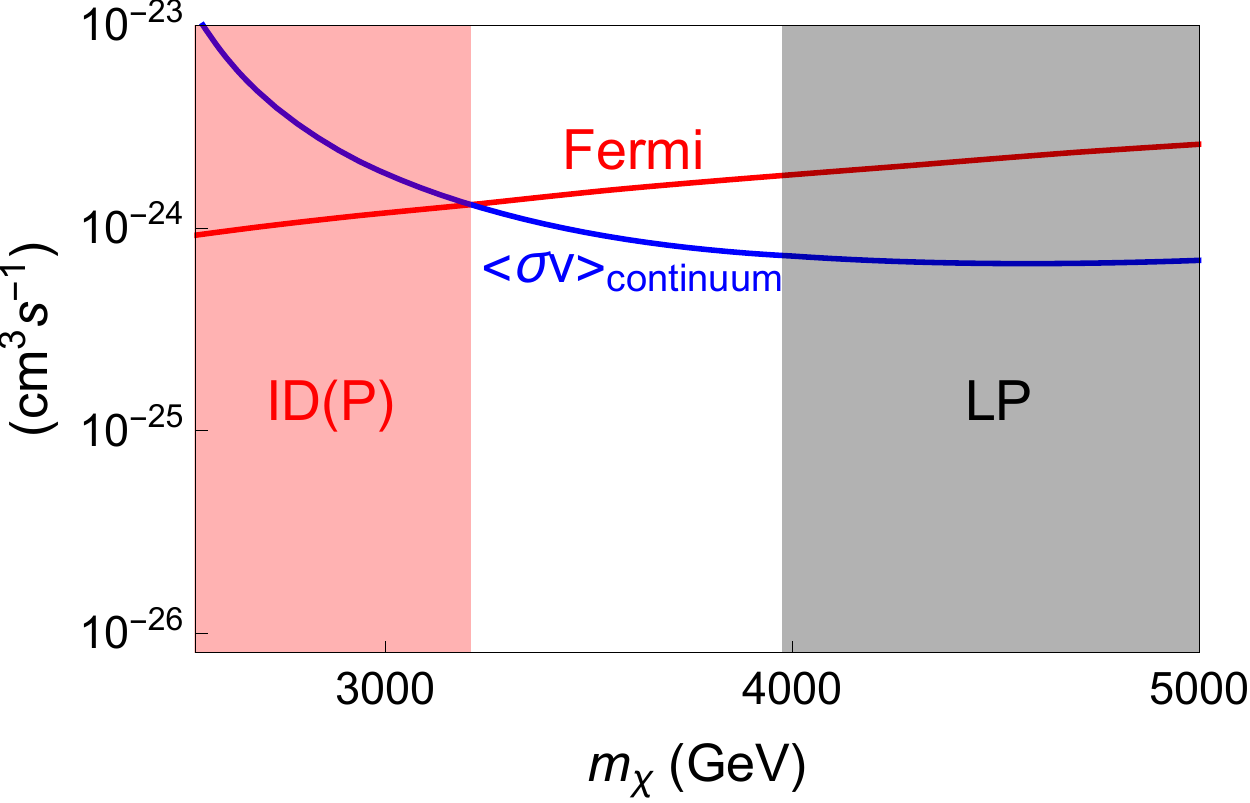}
    \quad
    \includegraphics[keepaspectratio, scale=0.61]{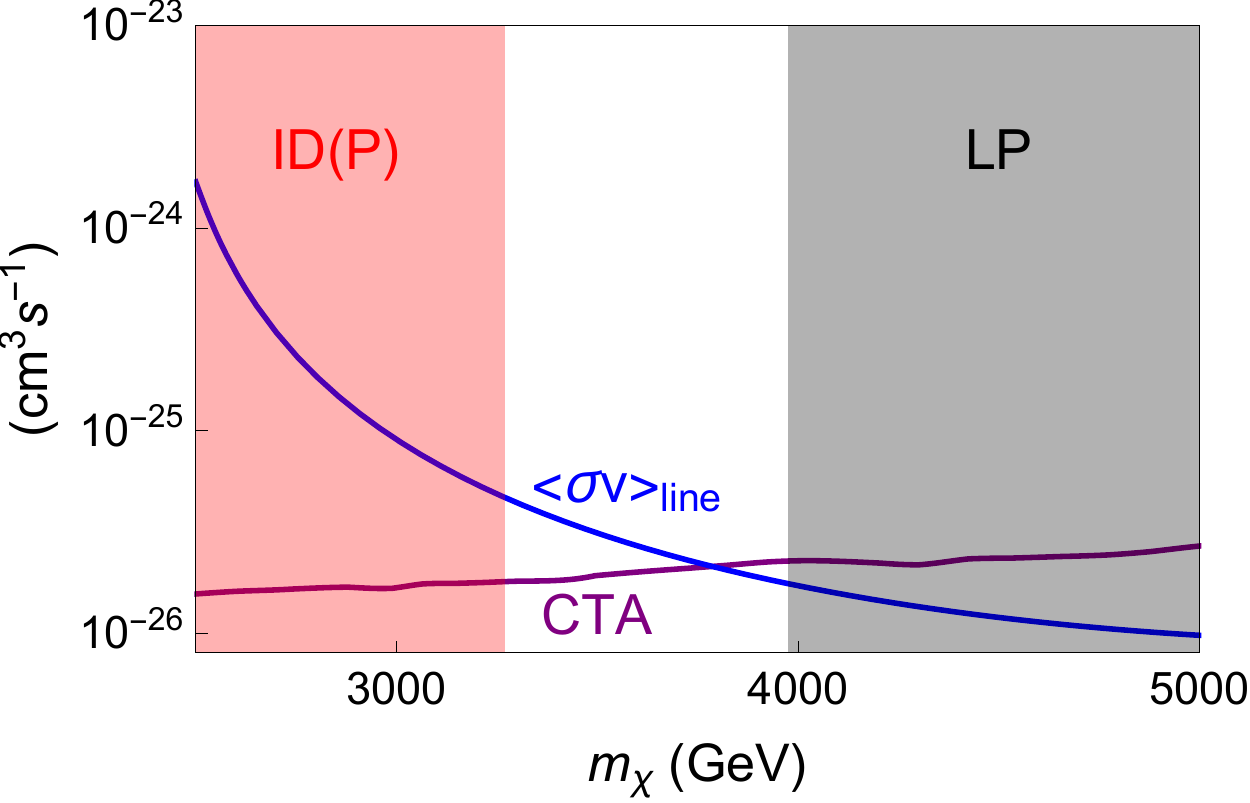}
    \caption{\small \sl The annihilation cross section producing the continuum gamma-ray signal and that producing the line gamma-ray signal are depicted by a blue line in left and right panels, respectively, as a function of the dark matter mass. The cross sections are compared with the present constraint from the Fermi-LAT experiment (red line) and the expected sensitivity of the CTA experiment (purple line), where the region above the lines are excluded/covered by the experiments. Here, the coupling $\lambda_{\chi H}$ is fixed at each dark matter mass $m_\chi$ so that it gives the thermal relic abundance which coincides with the dark matter density observed today. Shaded region below $\sim 3$\,TeV is forbidden because of the present constraint by the Fermi-LAT experiment, while that above $\sim 4$\,TeV is not preferred when we postulate that the dark matter theory does not break down up to high enough energy scale.}
    \label{indirectfig}
\end{figure}

Finally, the scalar triplet dark matter will also be efficiently searched for by the indirect dark matter detection observing dSph(s) at the CTA experiment, as also seen in Fig.\,\ref{constraintfig}. Importantly, when we focus on the model parameter region of $\lambda_{\chi H} \leq 0.5$, the wide parameter region of the `WIMP-Miracle' will be covered by this near future experiment as shown in Fig.\,\ref{indirectfig}, where the annihilation cross section producing the continuum gamma-ray signal and that producing the line gamma-ray signal are depicted as a function of the dark matter mass in left and right panels, respectively. Here, $\lambda_{\chi H}$ is fixed at a given $m_\chi$  so that it gives the thermal relic abundance coincides with the dark matter density observed today.

We have so far considered the simplest scalar triplet dark matter theory assuming that no further new physics does not affect physics of the dark matter (except the range of the coupling $\lambda_{\chi H}$). On the other hand, when such a new physics exists, it could significantly affect the dark matter physics via altering the mass difference between $\chi^0$ and $\chi^\pm$, even if its energy scale is not very close to the scale of the dark matter.\footnote{Such a new physics effect is indeed possible to be described by a higher-dimensional interaction, $|H^\dagger \chi H|^2$. In order to take the effect into account, the interaction should be added to the lagrangian defined in eq.\,(\ref{lagrangian}).} The constraint from collider experiments is then drastically weakened, and considering the scalar triplet dark matter with the mass $m_\chi \ll {\cal O}(1)$\,TeV becomes possible, though it is hard to realize the `WIMP-Miracle' scenario in such a case. This possibility has been thoroughly studied in Ref.\,\cite{Arakawa:2021vih}.

\appendix

\section{Beta functions}
\label{appendix: beta function}

Beta functions of the SM couplings, namely SU(3)$_c$ gauge coupling $g_s$, SU(2)$_L$ gauge coupling $g_2$, U(1)$_Y$ gauge coupling $g_Y$, top Yukawa coupling $y_t$ and Higgs self-coupling $\lambda$, at two-loop level, which are used to draw the lines in Fig.\,\ref{LPfig}, are given as follows:
\begin{align}
    \beta_X^{\rm (SM)} = \frac{1}{(4\pi)^2} \beta_X^{\rm (SM,1)} + \frac{1}{(4\pi)^4} \beta_X^{\rm (SM,2)},
\end{align}
\begin{align}
    &\beta_{g_Y}^{\rm (SM,1)} = \frac{41}{6} g_Y^3,
    \qquad\quad
    \beta_{g_Y}^{\rm (SM,2)} = g_{Y}^{3} \Big( \frac{199}{18} g_{Y}^{2}+\frac{9}{2} g_{2}^{2}  + \frac{44}{3} g_{s}^{2}  -\frac{17}{6}y_{t}^2 \Big), \\
    &\beta_{g_2}^{\rm (SM,1)} =  -\frac{19}{6} g_2^3,
    \qquad\,\,
    \beta_{g_2}^{\rm (SM,2)} = g_{2}^{3} \Big( \frac{3}{2} g_{Y}^{2} + \frac{35}{6} g_{2}^{2} + 12 g_{s}^{2}   -\frac{3}{2}y_{t}^2 \Big),\nonumber\\ 
    &\beta_{g_s}^{\rm (SM,1)} = -7 g_s^3,
    \qquad\quad\,
    \beta_{g_s}^{\rm (SM,2)} = g_{s}^{3} \Big(\frac{11}{6} g_{Y}^{2}   +\frac{9}{2}g_{2}^{2}  - 26 g_{s}^{2}  - 2 y_{t}^2 \Big), \nonumber
\end{align}
\begin{align}
    &\beta_{y_t}^{\rm (SM,1)} = y_t \left(\frac{9}{2} y_t^2 - \frac{17}{12} g_Y^2 - \frac{9}{4} g_2^2 - 8 g_s^2 \right), \\
    &\beta_{y_t}^{\rm (SM,2)} = -12y_{t}^5+y_{t}^3\Big(\frac{131}{16} g_{Y}^{2}  + \frac{225}{16} g_{2}^{2}+36 g_{s}^{2} -12 \lambda \Big) \nonumber \\ 
    &\qquad\qquad +y_{t} \Big(\frac{1187}{216} g_{Y}^{4} -\frac{3}{4} g_{Y}^{2} g_{2}^{2} -\frac{23}{4} g_{2}^{4} +\frac{19}{9} g_{Y}^{2} g_{s}^{2} +9 g_{2}^{2} g_{s}^{2} -108 g_{s}^{4} +6 \lambda^{2} \Big), \nonumber
\end{align}
\begin{align}
    &\beta_{\lambda}^{\rm (SM,1)} = 
    24 \lambda^2 - 6 y_t^4 + \frac{3}{8} \left( g_Y^4 + 2 g_Y^2 g_2^2 + 3 g_2^4 \right) + \lambda \left( 12 y_t^2 - 3 g_y^2 - 9 g_2^2 \right), \\
    &\beta_{\lambda}^{\rm (SM,2)} =-312 \lambda^{3}+ 36\lambda^{2}\Big( g_{Y}^{2} -4 y_{t}^2+3 g_{2}^{2} \Big)\nonumber \\  
    &\qquad\qquad +\lambda\Big(\frac{629}{24} g_{Y}^{4}  +\frac{39}{4} g_{Y}^{2} g_{2}^{2} -\frac{73}{8} g_{2}^{4}+y_{t}^2 \Big(\frac{85}{6} g_{Y}^{2} +\frac{45}{2} g_{2}^{2} +80 g_{s}^{2} \Big)\Big)\nonumber \\ 
    &\qquad\qquad+30 y_{t}^6- y_{t}^4\Big(\frac{8}{3} g_{Y}^{2} +32 g_{s}^{2} +3 \lambda\Big)-y_{t}^2\Big(\frac{19}{4} g_{Y}^{4}-\frac{21}{2} g_{Y}^{2} g_{2}^{2}+\frac{9}{4} g_{2}^{4}\Big)\nonumber \\ 
    &\qquad\qquad-\frac{379}{48} g_{Y}^{6} -\frac{559}{48} g_{Y}^{4} g_{2}^{2} -\frac{289}{48} g_{Y}^{2} g_{2}^{4} +\frac{305}{16} g_{2}^{6}. \nonumber
    \label{betaSM} 
\end{align}
Two-loop beta function including the contribution from a real scalar triplet (in addition to SM ones) are obtained using the SARAH code\,\cite{Staub:2013tta}, and those are given as follows:\footnote{The two-loop beta functions in the real triplet scalar theory can also be found in Refs.\,\cite{Khan:2016sxm,Jangid:2020qgo}.}
\begin{align}
    &\beta_{g_Y}^{(1)} = \beta_{g_Y}^{\rm (SM,1)},
    \qquad\qquad\quad
    \beta_{g_Y}^{(2)} = \beta_{g_Y}^{\rm (SM,2)}, \\
    &\beta_{g_2}^{(1)} = \beta_{g_2}^{\rm (SM,1)} + \frac{1}{3} g_2^3,
    \qquad
    \beta_{g_2}^{(2)} = \beta_{g_2}^{\rm (SM,2)} + \frac{28}{3} g_2^5, \nonumber \\
    &\beta_{g_s}^{(1)} = \beta_{g_s}^{\rm (SM,1)},
    \qquad\qquad\quad
    \beta_{g_s}^{(2)} = \beta_{g_s}^{\rm (SM,2)}, \nonumber
\end{align}
\begin{align}
    &\beta_{y_t}^{(1)} = \beta_{y_t}^{\rm (SM,1)},
    \qquad\qquad\quad
    \beta_{y_t}^{(2)} = \beta_{y_t}^{\rm (SM,2)} + y_t(g_{2}^4+3\lambda_{\chi H}^2),
\end{align}
\begin{align}
    &\beta_{\lambda}^{(1)} = \beta_{\lambda}^{\rm (SM,1)} + 6 \lambda_{\chi H}^2, \\
    &\beta_{\lambda}^{(2)} = \beta_{\lambda}^{\rm (SM,2)} + \lambda\left( \frac{11}{2}g_2^4 - 60 \lambda_{\chi H}^2 \right) -\frac{7}{12}g_Y^2g_2^4 - \frac{7}{4}g_2^6 + 30g_2^4\lambda_{\chi H} + 96 g_2^2\lambda_{\chi H}^2 - 48\lambda_{\chi H}^3, \nonumber
\end{align}
\begin{align}
    &\beta_{\lambda_\chi}^{(1)} = 88\lambda_\chi^2 + 3g_2^4 - 24\lambda_\chi g_2^2 + 2\lambda_{\chi H}^2, \\
    &\beta_{\lambda_\chi}^{(2)} = -4416 \lambda_{\chi}^{3}+1024 g_{2}^{2} \lambda_{\chi}^{2} +\lambda_\chi\Big(\frac{382}{3} g_{2}^{4} -80 \lambda_{\chi H}^{2}\Big)\nonumber \\
    &\qquad\quad -16 \lambda_{\chi H}^{3}+4\lambda_{\chi H}^{2}\Big(g_{Y}^{2} +3 g_{2}^{2} -3 y_{t}^2  \Big)+10 g_{2}^{4} \lambda_{\chi H}  -\frac{55}{3} g_{2}^{6}, \nonumber
\end{align}
\begin{align}
    &\beta_{\lambda_{\chi H}}^{(1)} = 
    3g_2^4 + 12\lambda \lambda_{\chi H}+40\lambda_\chi\lambda_{\chi H} + 6y_t^2\lambda_{\chi H} - \frac{3}{2}g_Y^2\lambda_{\chi H} - \frac{33}{2}g_2^2\lambda_{\chi H} + 8\lambda_{\chi H}^2, \\
    &\beta_{\lambda_{\chi H}}^{(2)} = 46 \lambda_{\chi H}^{3} + \lambda_{\chi H}^{2}\Big(2 g_{Y}^{2} +22 g_{2}^{2} -144 \lambda -480 \lambda_\chi-24 y_{t}^2 \Big)\nonumber \\ 
    &\qquad\quad+\lambda_{\chi H}\Big(\frac{557}{48} g_{Y}^{4} +\frac{15}{8} g_{Y}^{2} g_{2}^{2} +\frac{641}{48} g_{2}^{4} +24 g_{Y}^{2} \lambda  +72 g_{2}^{2} \lambda+640 g_{2}^{2} \lambda_\chi  \nonumber \\ 
    &\qquad\quad-\frac{27}{2}y_{t}^4 +y_{t}^2\Big(\frac{85}{12} g_{Y}^{2}  +\frac{45}{4} g_{2}^{2} +40 g_{s}^{2} -72 \lambda \Big)  -800\lambda_{\chi}^{2}   -60 \lambda^{2} \Big)\nonumber \\ 
    &\qquad\quad-\frac{15}{4} g_{Y}^{2} g_{2}^{4} +\frac{329}{12} g_{2}^{6} +30 g_{2}^{4} \lambda+100 g_{2}^{4} \lambda_\chi -3 g_{2}^{4} y_{t}^2. \nonumber
\end{align}

\section*{Acknowledgments}

This work is supported by Grant-in-Aid for Scientific Research from the Ministry of Education, Culture, Sports, Science, and Technology (MEXT), Japan; 17H02878 and 20H01895 (S.M. and S.S.), 19H05810 and 20H00153 (S.M.) and 18K13535, 20H05860 and 21H00067 (S.S.), by World Premier International Research Center Initiative (WPI), MEXT, Japan (Kavli IPMU), and also by JSPS Core-to-Core Program (JPJSCCA20200002).

\bibliographystyle{unsrt}
\bibliography{refs}

\end{document}